\documentclass{aa_gaia}

\usepackage{hyperref}

\let\orgautoref\autoref 
\renewcommand{\autoref}
        {\def\equationautorefname{Eq.}%
         \def\figureautorefname{Fig.}%
         \def\sectionautorefname{Sect.}%
         \def\subsectionautorefname{Sect.}%
         \def\subsubsectionautorefname{Sect.}%
         \orgautoref}
\newcommand{\Autoref}  % Capital at sentence start.
        {\def\equationautorefname{Equation}%
         \def\figureautorefname{Figure}%
         \def\sectionautorefname{Section}%
         \def\subsectionautorefname{Section}%
         \def\subsubsectionautorefname{Section}%
         \orgautoref}

\usepackage{xcolor}
\definecolor{dark-red}{rgb}{0.9,0.0,0.0}
\definecolor{dark-blue}{rgb}{0.15,0.15,0.9}
\definecolor{dark-green}{rgb}{0.15,0.8,0.15}
\definecolor{medium-blue}{rgb}{0,0,0.9}
\hypersetup{
    colorlinks, linkcolor=red,
    citecolor={dark-blue} , urlcolor={medium-blue}
}
\usepackage{changepage} % for \begin{adjustwidth}{-Xcm}{} \end{adjustwidth}

\makeatletter
\renewcommand*\aa@pageof{, page \thepage{} of \pageref*{LastPage}}
\makeatother

\usepackage{graphicx}
\usepackage{txfonts}
\usepackage{soul}
\usepackage{relsize}
\begin{document}

   \title{The CARMENES search for exoplanets around M dwarfs}

%  and prospects of habitability 

  % \subtitle{Dynamical characterization of the multiple planet system Ross 1003}
    \subtitle{
     Dynamical characterization of the multiple planet system GJ\,1148
     and prospects of habitable exomoons around GJ\,1148 b}

     \author{T.\,Trifonov\inst{1}
     \and Man\,Hoi\,Lee \inst{2,3}     
     \and M.\,K\"urster\inst{1}  
     \and Th.\,Henning\inst{1} 
     \and E.\, Grishin\inst{4}            
     \and S.\,Stock\inst{5}   
     \and J.\,Tjoa\inst{6}       
     \and J.\,A.\,Caballero\inst{7}     
     \and Ka\,Ho\,Wong\inst{2}    
     \and F.\,F.\,Bauer\inst{8}
     \and A.\,Quirrenbach\inst{5}     
     \and M.\,Zechmeister\inst{9}     
     \and I.\,Ribas\inst{10,11}
     \and S.\,Reffert\inst{5}     
     \and A.\,Reiners\inst{9}      
     \and P.\,J.\,Amado\inst{8} 
     \and D.\,Kossakowski\inst{1}         
     \and M.\,Azzaro\inst{12} 
     \and V.\,J.\,S.\,B\'ejar\inst{13,14}
     \and M.\,Cort\'es-Contreras\inst{7}
     \and S.\,Dreizler\inst{9}
     \and A.\,P.\,Hatzes\inst{15}
     \and S.\,V.\,Jeffers\inst{9}
     \and A.\,Kaminski\inst{5}
     \and M.\,Lafarga\inst{10,11}
     \and D.\,Montes\inst{16}    
     \and J.\,C.\,Morales\inst{10,11}
     \and A.\,Pavlov\inst{1}
     \and C.\,Rodr\'iguez-L\'opez\inst{8} 
     \and J.\,H.\,M.\,M.\,Schmitt\inst{17}
     \and E.\,Solano\inst{7}
     \and R. Barnes\inst{18,19}
}
% ----------- MPIA
   \institute{Max-Planck-Institut f\"ur Astronomie,
              K\"onigstuhl 17, D-69117 Heidelberg, Germany\\
              \email{trifonov@mpia.de}
% ----------- Hong Kong - Earth [Lee, Ho]
          \and Department of Earth Sciences, The University of Hong Kong, Pokfulam Road, Hong Kong
% ----------- Hong Kong - Physics [Lee]
          \and Department of Physics, The University of Hong Kong, Pokfulman Road, Hong Kong
% ----------- Technion [Grishin]
          \and Physics Department, Technion - Israel institute of Technology, Haifa, Israel 3200002
% ----------- LSW
         \and Landessternwarte, Zentrum f\"ur Astronomie der Universt\"at Heidelberg,
              K\"onigstuhl 12, D-69117 Heidelberg, Germany  
% ----------- Kapteyn [Tjoa]
          \and Kapteyn Astronomical Institute, University of Groningen, Landleven 12, 9747 AD  Groningen, The Netherlands
% ----------- CAB [Villafranca]
         \and Centro de Astrobiolog\'ia (CSIC-INTA), ESAC, camino bajo del castillo s/n, 28049 Villanueva de la Ca\~nada, Madrid, Spain
         % ----------- IAA
         \and Instituto de Astrof\'isica de Andaluc\'ia (IAA-CSIC), Glorieta de la Astronom\'ia s/n, 
              E-18008 Granada, Spain
% ----------- IAG
          \and Institut f\"ur Astrophysik, Georg-August-Universit\"at, 
              Friedrich-Hund-Platz 1, 37077 G\"ottingen, Germany              
% ----------- ICE-1
         \and Institut de Ci\`encies de l'Espai (ICE, CSIC), Campus UAB, c/ de Can Magrans s/n, 
              E-08193 Bellaterra, Barcelona, Spain
% ----------- ICE-2
         \and Institut d'Estudis Espacials de Catalunya (IEEC), C/ Gran Capit\`a 2-4, 08034 Barcelona, Spain
% ----------- CAHA
          \and Centro Astron\'omico Hispano-Alem\'an (CSIC-MPG), 
               Observatorio Astron\'omico de Calar Alto, 
               Sierra de los Filabres, E-04550 G\'ergal, Almer\'ia, Spain
% ----------- IAC
          \and Instituto de Astrof\'sica de Canarias, V\'ia L\'actea s/n, 38205 La Laguna, Tenerife, Spain 
          \and Departamento de Astrof\'isica, Universidad de La Laguna, 38206 La Laguna, Tenerife, Spain
% ----------- TLS
         \and Th\"uringer Landessternwarte Tautenburg, Sternwarte 5, D-07778 Tautenburg, Germany
% ----------- UCM
          %\and Departamento de Astrof\'isica y Ciencias de la Atm\'osfera, Facultad de Ciencias F\'isicas, Universidad Complutense de Madrid, E-28040 Madrid, Spain
          \and Departamento de F\'{i}sica de la Tierra y Astrof\'{i}sica 
and IPARCOS-UCM (Intituto de F\'{i}sica de Part\'{i}culas y del Cosmos de la UCM), 
Facultad de Ciencias F\'{i}sicas, Universidad Complutense de Madrid, E-28040, Madrid, Spain
% ----------- HS
         \and Hamburger Sternwarte, Gojenbergsweg 112, D-21029 Hamburg, Germany
         \and Astronomy Department, University of Washington, Seattle, WA, USA 98195
         \and NASA Virtual Planetary Laboratory
              }

   \date{Received 25 October 2019 / Accepted 28 January 2020}
 
% \abstract{}{}{}{}{} 
% 5 {} token are mandatory

  \abstract
   {GJ\,1148 is an M-dwarf star hosting a planetary system composed of two Saturn-mass planets in eccentric orbits with periods of 41.38 and 532.02 days. }   
%   % aims heading (mandatory)
   {
  We reanalyze the orbital configuration and dynamics of the GJ\,1148 multi-planetary system based on new
  precise radial velocity (RV) measurements taken with CARMENES. %HIRES and CARMENES.
% for these planet hosts and test the overall capabilities of CARMENES.  
   }
   {
    We combined new and archival precise Doppler measurements from CARMENES with those available from HIRES for 
    GJ\,1148 and 
    modeled these data with a self-consistent dynamical model. We studied the 
    orbital dynamics of the system using the secular theory and direct N-body integrations.
    The prospects of potentially habitable moons around GJ\,1148 b 
    were examined. 
   }
   {
  The refined dynamical analyses show that the GJ\,1148 system is long-term stable in a large phase-space of orbital parameters with an orbital configuration suggesting apsidal alignment, but not in any particular high-order mean-motion resonant commensurability.
   GJ\,1148 b orbits inside the optimistic habitable zone (HZ).
   We find only a narrow stability region around the planet where exomoons can exist. 
  However, in this stable region exomoons  exhibit quick orbital decay due to tidal interaction with the planet.
   }
   {
  The GJ\,1148 planetary system is a very rare M-dwarf planetary system consisting of a pair of gas giants, the inner of which resides in the HZ. 
    We conclude that habitable exomoons around GJ\,1148 b are very unlikely to exist.
   }

   \keywords{planetary systems -- optical: stars -- stars: late-type -- stars: low-mass -- planets and satellites: dynamical evolution and stability}
   
   \authorrunning{Trifonov et al.}
   \titlerunning{Dynamical characterization of GJ\,1148}

   \maketitle

%________________________________________________________________

\section{Introduction}

In the past twenty years, M-dwarf stars have been the primary targets for a number of planet search surveys
via the precise Doppler method \citep{Marcy1998,Delfosse1998,Marcy2001,Endl2003,Kurster2003,Bonfils2005,Butler2006,Zechmeister2009b,Reiners2018a}. 
These efforts are motivated by the fact that M dwarfs are the predominant population of stars in the solar neighborhood. 
Their significantly 
lower mass and luminosity when compared to Sun-like stars 
make them suitable for the detection of lower mass planets orbiting in the so-called 
habitable zone (HZ), where the surface temperature would be favorable for liquid water to exist.
Planets in the HZ have already been discovered around low-mass 
M dwarf neighbors such as Proxima Centauri \citep{Anglada2016}, 
LHS 1140 \citep{Dittmann2017}, HD\,147379 \citep{Reiners2018b}, GJ\,752A \citep{Kaminski2018},
and GJ\,357 \citep{Luque2019}, among others.
 Of special interest for Doppler surveys are also potentially habitable M-dwarf multiple planet systems 
consisting of Earth-size planets such as GJ\,1069 \citep{Dreizler2019}, or the ultra-cool M~dwarfs TRAPPIST-1 \citep{Gillon2017} and Teegarden's star \citep{Zechmeister2019}.

\begin{table} 
\caption{Physical parameters information of GJ\,1148} %
\label{table.stars}
\centering
\begin{tabular}{@{}l l r@{}}
\hline
\hline
\noalign{\smallskip}
 \hspace{35mm}                            & \object{GJ\,1148}         & Ref. \\
\hline
\noalign{\smallskip}
Karmn                         & J11417+427                 &  Cab16      \\
Simbad                        & Ross\,1003                 &   Ros39      \\
HIP                           & 57050                      &   HIP     \\
%\noalign{\smallskip}
Sp. type                      & M4.0\,V                    & Rei18     \\
$G$ [mag]                     & 10.5769$\pm$0.0006         & {\em Gaia}     \\
$J$ [mag]                     & 7.608$\pm$0.018            & 2MASS     \\
$d$ [pc]                      & 11.02$\pm$0.01             & {\em Gaia} \\
$\mu_\alpha \cos{\delta}$ [mas\,a$^{-1}$]       & --575.65$\pm$0.07     & {\em Gaia} \\
$\mu_\delta$ [mas\,a$^{-1}$]  & --89.973$\pm$0.07   & {\em Gaia} \\
%$V_r$ [km\,s$^{-1}$]          & +35.678                    & Rei18        \\
%$U$ [km\,s$^{-1}$]            & +53.2         & This work     \\
%$V$ [km\,s$^{-1}$]            & --7.6         & This work     \\
%$W$ [km\,s$^{-1}$]            & --5.0         & This work     \\
$v \sin{i}$ [km\,s$^{-1}$]    & $<$2 & Rei18        \\
$T_{\rm eff}$ [K]             & 3358$\pm$51                & Pas18        \\
$\log{g}$                     & 4.99$\pm$0.07              & Pas18        \\
{[Fe/H]}                      & 0.13$\pm$0.16              & Pas18        \\
$L$ [$L_\odot$]               & 0.0143$\pm$0.0003          & Schw19     \\
$R$ [$R_\odot$]               & 0.353$\pm$0.014            & Schw19    \\
$M_\star$ [$M_\odot$]         & 0.354 $\pm$ 0.015          & Schw19 \\
%pEW(H$\alpha$) [{\AA}]        & +0.3$\pm$0.1 & Jef18 \\
%$\log{R'_\mathrm{HK}}$        & -5.071$\pm$0.071 & Ast17\\
$P_{\rm rot}$ [d]             & 71.5$\pm$5.1               & DA18\\
\noalign{\smallskip}
\hline
\end{tabular}
% $$
\tablefoot{\small 
2MASS: \citet{Skrutskie2006};
Cab16 \citet{Caballero2016};
DA18: \citet{DiezAlonso2019};
{\em Gaia} \citet[Gaia DR2;][]{Gaia_Collaboration2018b}; % Gaia_Collaboration2016
Pas18: \citet{Passegger2018};
 HIP: \citet{Perryman1997};
Rei18: \citet{Reiners2018a};
Ros39: \citet{Ross1939};
Schw19: \citet{Schweitzer2019}.
}
\end{table}

As of October 2019,  more than 100 M-dwarf planetary systems have been discovered via the radial velocity (RV) method \citep{MartinezRodriguez2019}\footnote{See also \url{http://exoplanet.eu}}, 
Many more are expected to be detected with the transit technique thanks to the 
Transiting Exoplanet Survey Satellite \citep[TESS;][]{Ricker2015}, which already
delivered the first planet detections around M-dwarf stars 
such as \object{L~95--98} \citep{Kostov2019,Cloutier2019}, TOI~270 \citep[LEHPM~3808;][]{Gunther2019}, and \object{GJ~357} \citep{Luque2019},
which are in fact all multi-planet systems.

The recent bulk of multi-planetary discoveries around M dwarfs opens a
great opportunity to study their orbital architecture in detail. 
Understanding the dynamics of multi-planet systems around M dwarfs provides an important clue to planet formation around these low-mass stars 
\citep{Lissauer2007,Raymond2007,Zhu2012,AngladaEscude2013,Coleman2017,Morales2019}. Furthermore, the planetary dynamics affect the evolution of any natural satellites; these exomoons can provide critical information on the formation and evolution of a planetary system \citep[see][]{Heller2014}, and may even be habitable themselves \citep{Williams1997,HellerBarnes2013,Forgan2016}. While no exomoon detection has been confirmed  \citep[see, e.g.,][]{Teachey2018,Heller2019}, theoretical studies have shown that the planets' orbital oscillations can drive the orbital evolution of satellites \citep{Zollinger2017} and that the tidal evolution of the satellite's orbit about the host planet can also be significant \citep[e.g.,][]{MartinezRodriguez2019}.

In this paper, we provide the first detailed analysis of the dynamics of the
rare GJ\,1148 system, which was reported to host two Saturn-mass planets 
orbiting around the M dwarf on moderately eccentric orbits \citep{Trifonov2018a}.
We provide new precise Doppler measurements from the CARMENES\footnote{Calar Alto high-Resolution search for
M dwarfs with Exo-earths with Near-infrared and optical Echelle Spectrographs, \url{http://carmenes.caha.es}} 
instrument \citep{Quirrenbach2016, Reiners2018a}
and focus on the long-term dynamical properties of the system, which could provide important information on the primordial protoplanetary conditions 
needed to form two Saturn-mass planets around a low-mass star.

The paper is  organized  as follows. 
In \autoref{Sec2} we introduce the updated physical properties of the M-dwarf star GJ\,1148 and provide a short summary of the literature on the known planetary system. In \autoref{Sec3} we present the observational data, and in \autoref{Sec4} we present our data analysis.
In \autoref{Sec5} we detail our dynamical results of the GJ\,1148 system. In \autoref{Sec6} we provide our conclusions and a brief discussion.

\begin{figure}
    \centering
    \includegraphics[width=9cm]{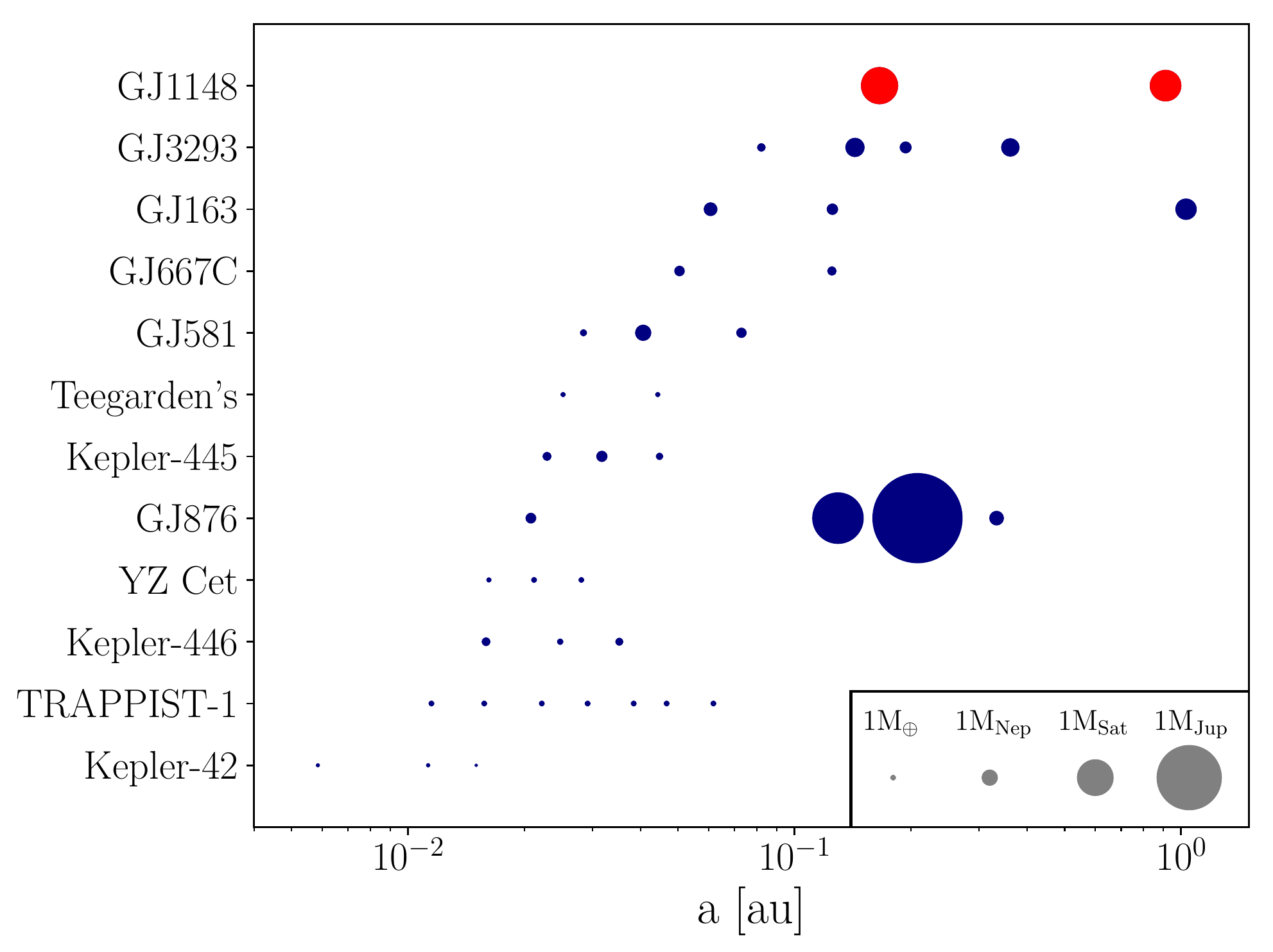}
    \caption{Architecture of known multi-planet systems around M-dwarf hosts, ordered by 
semi-major axis of the innermost planet (list is not complete).
Most of the M-dwarf planetary systems are composed of low-mass planets in or below the Neptune-mass
regime. Notable exceptions are the GJ\,876 and the GJ\,1148 M-dwarf systems, which are 
consistent with a pair of planets in the Jupiter- and Saturn-mass regimes, respectively.
     }
    \label{multiplanet_Mdwarfs} 
\end{figure}

\begin{figure*}
    \centering
    \includegraphics[width=9cm]{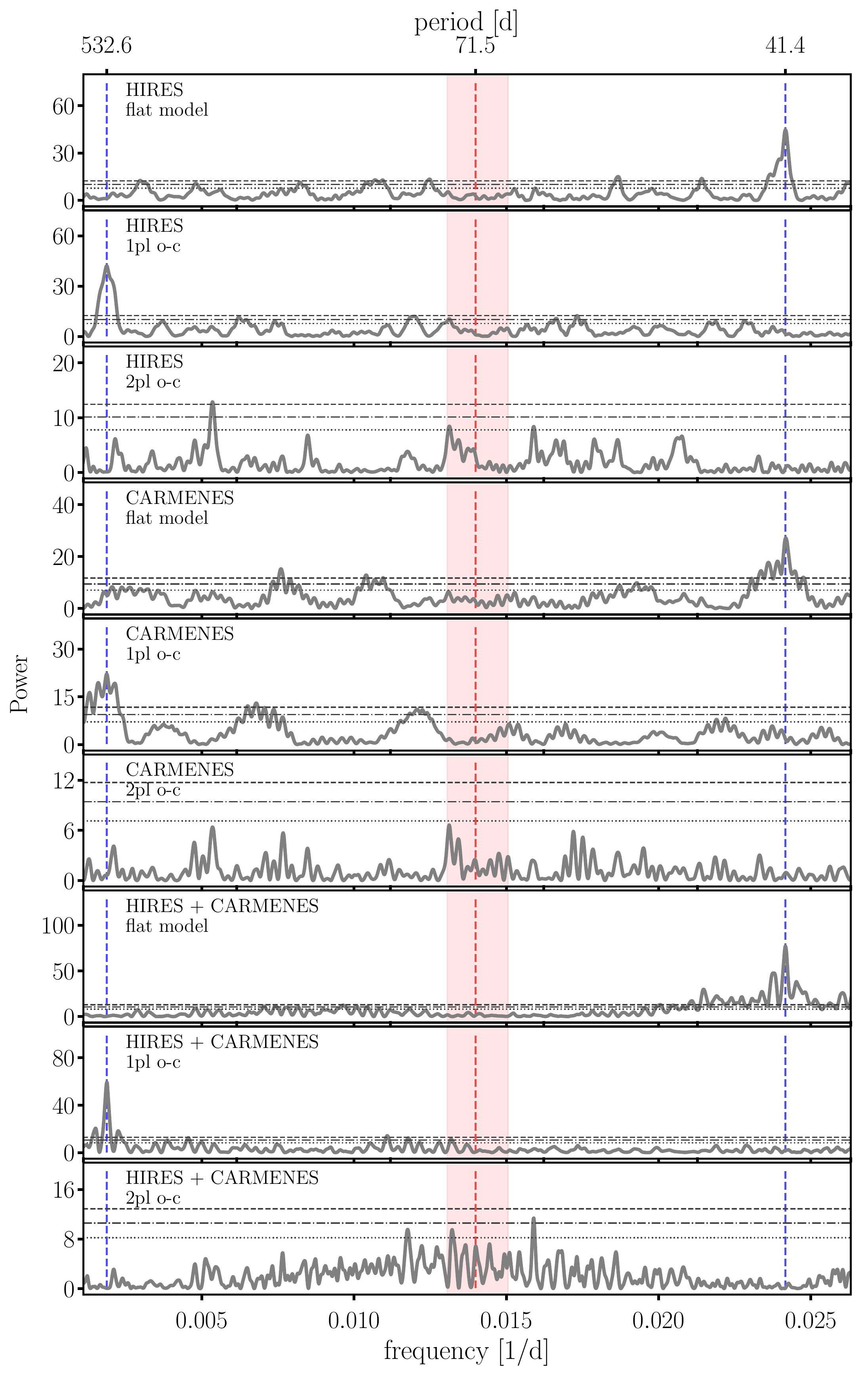} 
    \includegraphics[width=9cm]{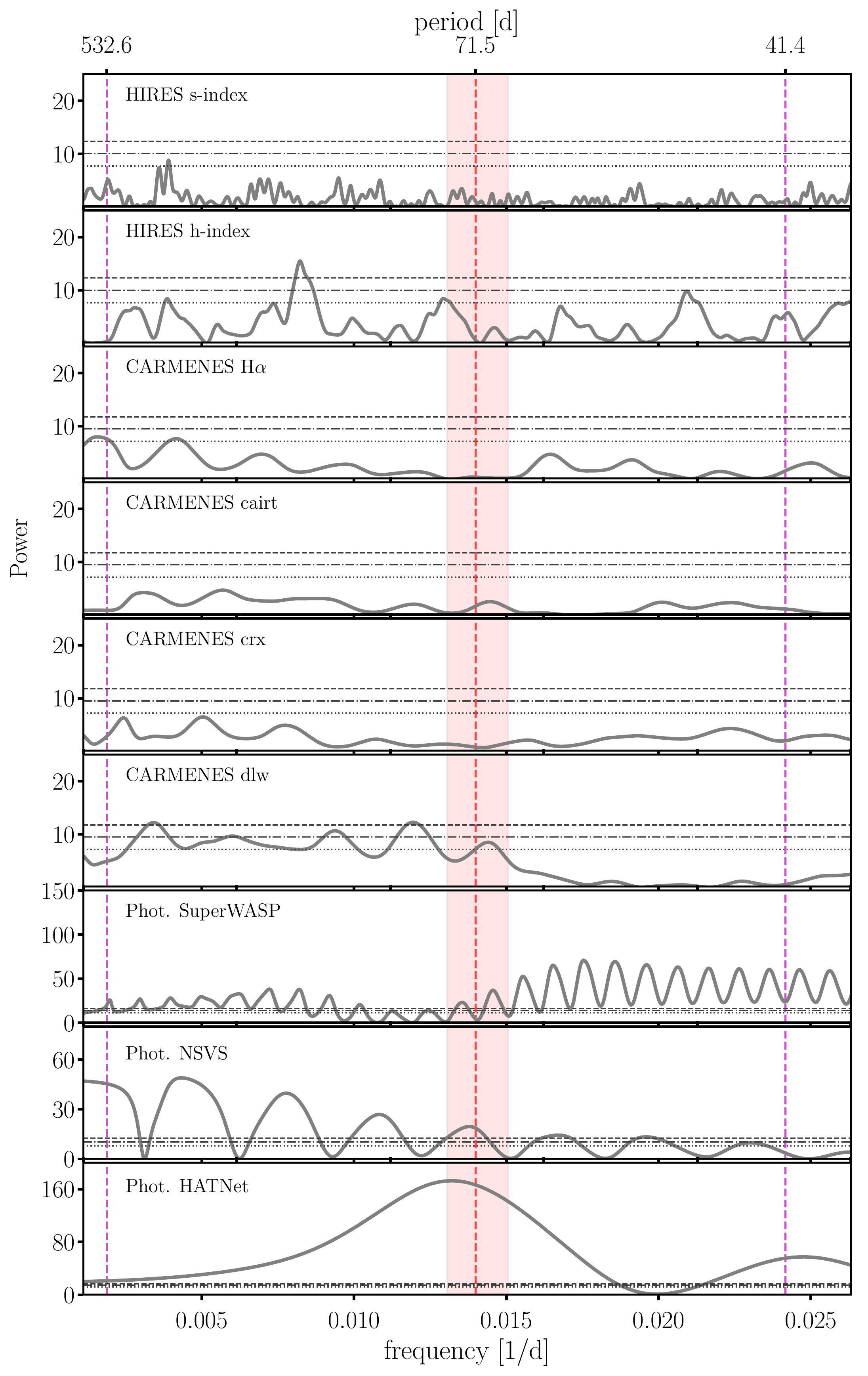} \\
    \caption{{\em Left panels}: GLS periodograms of the CARMENES and HIRES Doppler measurements 
combined and separately after consecutive  subtraction of the GJ\,1148 b and c planetary signals.
{\em Right panels}:  GLS peridograms of the CARMENES and HIRES activity indicators of GJ\,1148. 
Horizontal lines show  the FAP levels of 10\% (dotted line), 1\% (dot-dashed line), and 0.1\% (dashed line).
The S- and H-index measurements from the HIRES data and the CARMENES activity indicators have no power at the period of the planets GJ\,1148 b and c (vertical blue dashed lines). 
The photometric measurements from SuperWASP, NSVS, and HATNet are densely sampled, and thus most of the
GLS power appears significant, but it cannot be associated with the RV signals.
The most likely stellar rotation frequency is shown with red dashed line,
while the red shaded area denotes its uncertainties, as given by \citet{DiezAlonso2019}.
    }
    \label{s_h_index} 
\end{figure*}

\section{The GJ\,1148 planetary system}
\label{Sec2}

The planet GJ\,1148 b was discovered by \citet{Haghighipour2010} based on 37 velocities taken with the Keck/HIRES spectrograph \citep{Vogt1994}.
It is an eccentric Saturn with a period of $\sim$41.4\,d and, interestingly, mostly resides in the optimistic habitable zone (HZ).
After the public release of the HIRES velocity archive by \citet{Butler2017},
it became clear that the extended HIRES data time series yields an additional RV signal
with a period of $\sim$ 530\,d, likely due to a second massive planet in the system.
In \citet{Trifonov2018a}, we added 52 CARMENES RVs and performed a combined Doppler analysis,
which revealed the two-planet configuration of the GJ\,1148 system. We found the following orbital configuration for GJ\,1148~b:
$m_{\rm b} \sin i$  = 0.304\,$M_{\rm Jup}$,
$P_{\rm b}$ = 41.380\,d,
$e_{\rm b}$ = 0.379,
$a_{\rm b}$ = 0.166\,au,
and for GJ\,1148~c:  
$m_c \sin i$ = 0.214\,$M_{\rm Jup}$,
$P_{\rm c}$ = 532.6\,d,
$e_{\rm c}$ = 0.341,  
$a_{\rm c}$ = 0.913\,au.

\Autoref{multiplanet_Mdwarfs} 
shows the architectures of multi-planet systems known to orbit M dwarfs, including GJ 1148.
To date the observational results indicate that the planetary population around 
M dwarfs consists predominately of low-mass planets in the range 
from a few Earth masses to Neptune mass.
Together with the GJ\,876 system \citep{Rivera2010, Nelson2016, Millholland2018}
the GJ\,1148 system 
belongs to the rare population of M dwarfs known to host a pair of Jupiter-mass planets.
In contrast, observations of solar-mass main-sequence stars and 
more massive giant and sub-giant stars show a higher frequency of Jupiter-mass planets, which seems to be correlated with stellar host mass \citep{Fischer2005,Reffert2015}.
These findings suggest that more massive stars tend to have more massive planets, likely due to a more massive primordial protoplanetary disk  from which planets form.
In this context M dwarfs are not expected to have a large population of 
Jupiter-mass planets, which makes systems like GJ\,1148  an important 
discovery that may provide insights into the 
primordial disk properties around low-mass stars needed to accumulate gas giant planets.

\begin{figure*}
    \centering
    \includegraphics[width=\textwidth]{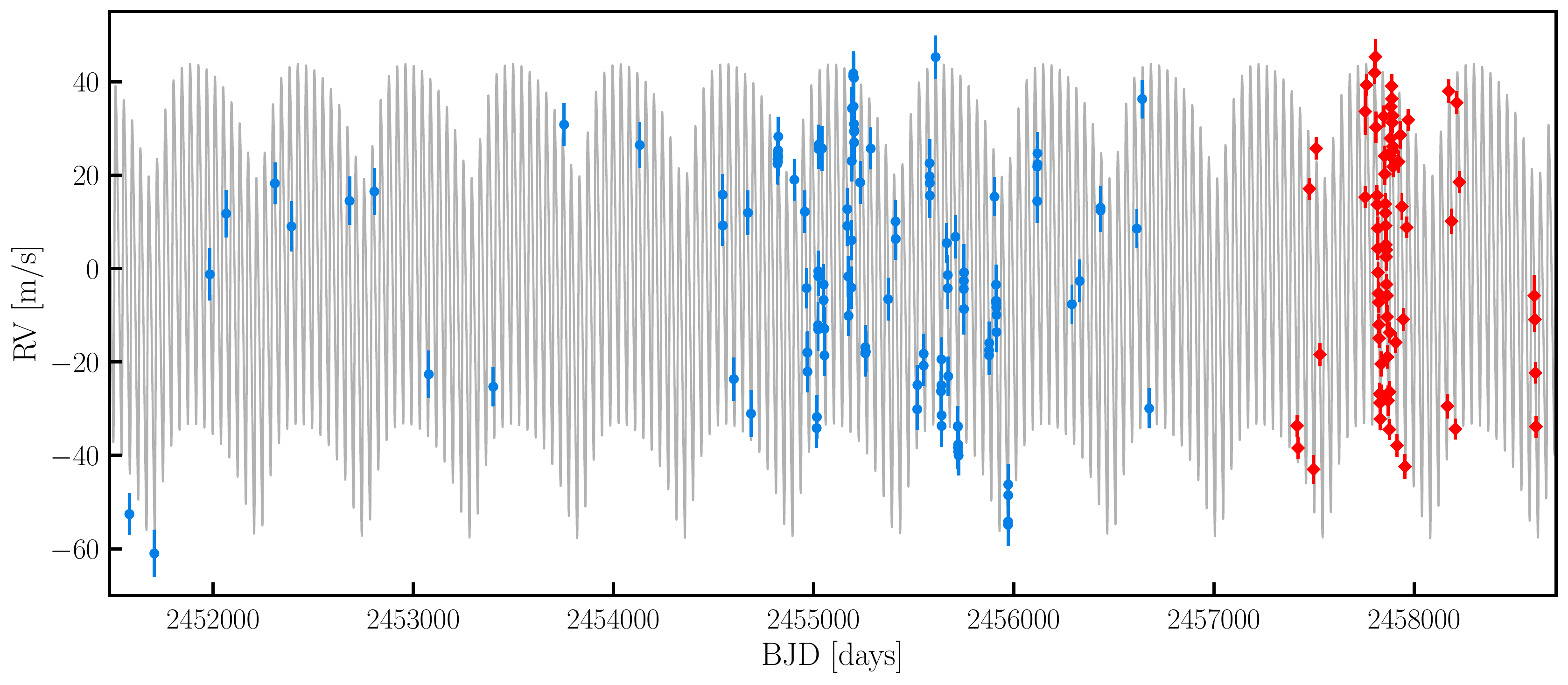} \\
    \includegraphics[width=\textwidth]{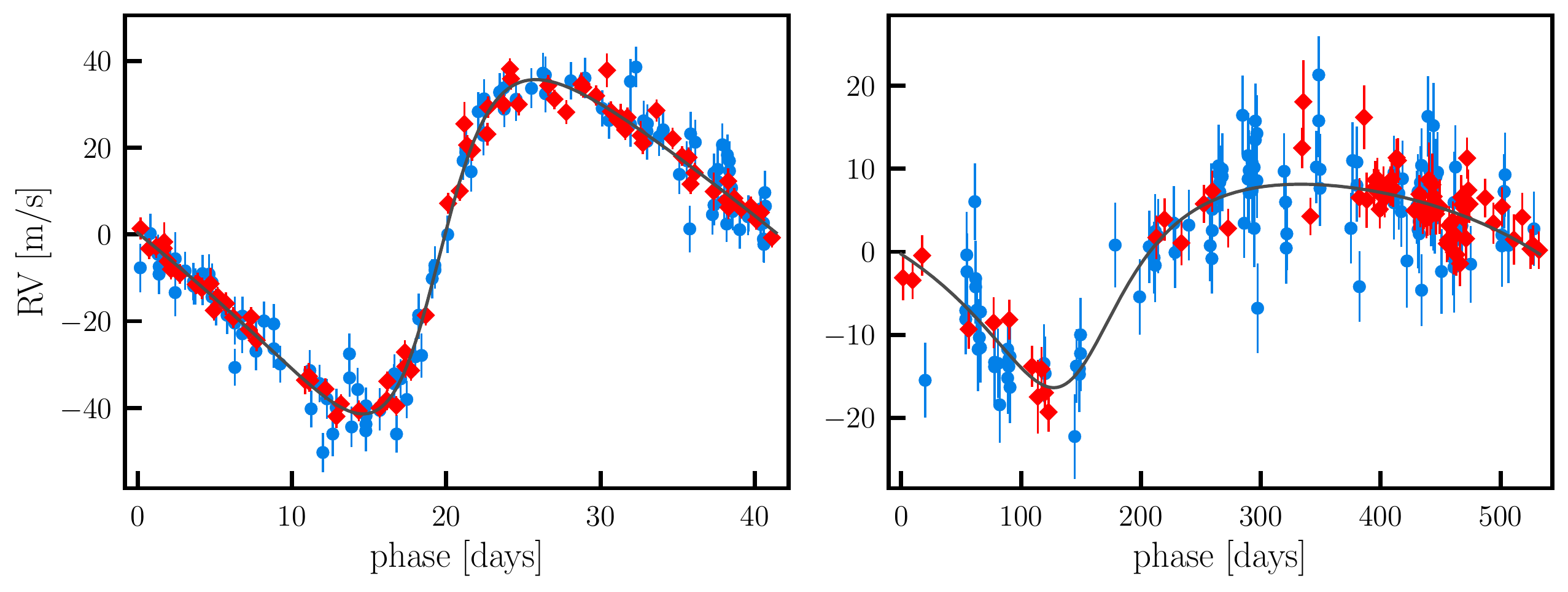}  \put(-450,170){GJ 1148 b} \put(-200,170){GJ 1148 c}
     
    \caption{
    {\em Top panel}: Precise optical Doppler measurements of GJ\,1148 from HIRES (blue triangles) and CARMENES (red circles) modeled with a two-planet model. 
    {\em Bottom panels}: Phase-folded planetary signals fitted to the data. Compared to \citet{Trifonov2018a}, %with the difference 
there are 24 additional CARMENES RVs, and the HIRES data were corrected for small systematics.  The data uncertainties include the estimated RV jitter.
}
\label{new_data} 
\end{figure*}

Apart of the GJ\,1148 and the GJ\,876 M-dwarf planetary systems, there are some other peculiar exceptions; 
for example, GJ\,1046 hosts 
a brown dwarf companion on 169-day period orbit  \citep{Kuerster2008,Trifonov2018c},
and there is an unexpected giant planet around the very low-mass M-dwarf star GJ\,1132 \citep{Morales2019}. The GJ\,317 system also has a Jupiter-mass planet in orbit \citep{Johnson2007}.
The observational evidence suggest that the GJ\,1132 and GJ\,317 M dwarfs 
are very likely accompanied by 
another long-period giant planet whose orbit has not yet been precisely determined.

In our first paper on GJ\,1148 \citep{Trifonov2018a}, we performed dynamical modeling of the GJ\,1148 RV data and we studied the long-term stability of the best fit to the system,
but at that time we did not perform a detailed long-term dynamical analysis
of the GJ\,1148 system. In this second paper, we aim to analyze the 
possible orbital configurations and planetary mass regimes based on an extensive long-term 
dynamical analysis that includes additional CARMENES Doppler data from recent observations.\looseness=-28

Based on analyses of CARMENES spectra taken between January 2016 and  January 2019, \citet{Schweitzer2019} and \citet{Passegger2018}
updated the physical parameters of the GJ\,1148 host star.
The results from these two studies report that GJ\,1148 is a typical 
M4.0V star with a stellar mass of 0.354 $\pm$ 0.015 $M_\odot$, effective temperature
$T_{\rm eff}$ = 3358$\pm$51\,K, and luminosity $L$ = 0.0143$\pm$0.0003 $L_\odot$.  
\autoref{table.stars} summarizes these 
new estimates for GJ\,1148 as well the recent estimates of other parameters from {\em Gaia} \citep[Gaia DR2;][]{Gaia_Collaboration2016, Gaia_Collaboration2018b}.

%%%%%%%% GJ1148 and TESS %%%%%%%%%%%%%%%%%%

% TIC     |   RA      |   Dec     | EclipticLong | EclipticLat | Sector | Camera | Ccd | ColPix | RowPix
% 115869504 | 175.435983 |  42.751974 | 156.148082 |  36.839007 | 22 | 2 | 3 | 1064.572 | 1266.955
% 
% Sector 22     from 02/18/20 to 03/18/20       orbits 51, 52
%  
 
% 18 Feb = 2458897.500000, 
% while the transit is likely to occur on JD = 2458892.983900, which is
% Feb 13 2020   11:36:49.0 UT
%  
% Next one will likely be 
% 
% Mar 25 20:44:01.0 UT, which means TESS will skip the potential transit of GJ 1148 b!
% 

% no chance for GJ1148 c. 

% prob of transit 

% In [1]: St_rad = 0.353 * 0.00465047
% 
% In [2]: a_b = 0.166
% 
% In [3]: e_b = 0.375
% 
% In [4]: ``Eq 8 in \citet{Barnes2007}
% 
% In [4]: prob = St_rad/ a_b*(1.0 - e_b**2)
% 
% In [5]: prob
% Out[5]: 0.008498576341302709

\begin{figure*}[tp]
\begin{center}$
    \begin{array}{cc} 
    \includegraphics[width=18cm]{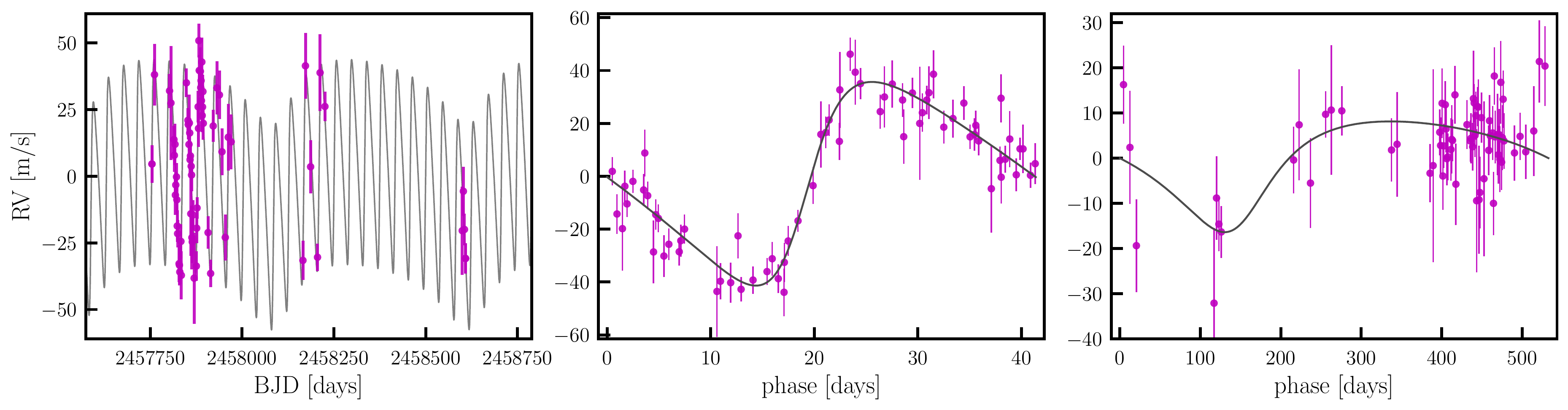} \put(-305,110){\tiny GJ 1148 b} \put(-135,110){\tiny GJ 1148 c} 
    \end{array} $

\end{center}
\caption{
From left to right: CARMENES near-IR Doppler measurements time series
of GJ\,1148 and a phase-folded representation of the two planetary signals similar to that of \autoref{new_data}.
Due to its  lower precision the near-IR data was not used in 
the data analysis, but is only overplotted with an optimized RV offset 
to the best two-planet dynamical model constructed from the optical data shown in \autoref{new_data}.
The CARMENES near-IR RVs are in an excellent agreement with the model from the available optical data, 
fully recovering the two planetary signals. 
}
\label{NIR_data} 
\end{figure*}

\section{Radial velocity data}
\label{Sec3}

\subsection{HIRES data}
\label{Sec3.1}

HIRES is a high-resolution spectrograph mounted on the 10 m Keck-I telescope, Keck Observatory, Hawaii, USA, \citep{Vogt1994}; via an iodine (I$_{\rm 2}$) cell \citep{Marcy1992} it can achieve RV precision down to $\sim$ 3\,ms$^{-1}$ \citep{Butler1996}.
\citet{Butler2017} have published 125 precise HIRES RV measurements of GJ\,1148 
as part of a large archive of precise Doppler measurements
and stellar-line activity-index measurements ($\sim 65\,000$ HIRES spectra of $\sim1700$ stars), obtained between 1996 and 2014.
Recently, the large sample size of the HIRES RV archive has allowed \citet{TalOr2018} 
to identify small but significant systematic nightly zero-point variations in the data on the order of $\sim$ 1\,ms$^{-1}$, and to correct for them,
making the HIRES data even more precise.

We note that for GJ\,1148 in particular, the small 
systematic corrections found in  \citet{TalOr2018} have 
a negligible effect since the planetary signals are significantly above the data noise level.
However, the data from \citet{TalOr2018} is also corrected for the small RV offset 
caused by the CCD upgrade performed in August 2004 \citep{Butler2006}, 
and thus, in this paper we 
decided to perform our analysis with the data provided by \citet{TalOr2018}.

\subsection{CARMENES data}
\label{Sec3.2}

A detailed description of the CARMENES instrument can be found in \citet{Quirrenbach2016},
while the CARMENES survey and goals are described in more detail in \citet{Reiners2018a}.
The standard CARMENES data reduction flow is described in \citet{Zechmeister2018}.
 
Briefly, the CARMENES instrument consists of two high-resolution spectrographs, which are 
designed to perform state-of-the-art RV measurements in the optical (0.52\,$\mu$m to 0.96\,$\mu$m) 
and in the  near-IR (CARMENES-NIR; 0.96\,$\mu$m to 1.71\,$\mu$m).
We obtained 76 optical and 68 near-IR CARMENES RV measurements of GJ\,1148 between January 2016 and May 2019. 
The uneven number of optical and  near-IR CARMENES data is due to 
NIR channel calibration issues (typically during the first six-months of the CARMENES operation).
For eight of the near-IR spectra we were unable to obtain meaningful RVs measurements, and thus
these data were discarded.
The first 52 optical RV measurements were published in \citet{Trifonov2018a}.
We computed all RVs with the latest pipeline and the SpEctrum Radial
Velocity AnaLyser  \citep[SERVAL;][]{Zechmeister2018} version, thus creating a revised and expanded data set. 
Therefore, we present 24 new and 52 reprocessed optical RVs and a new set of near-IR data of GJ\,1148.

Additionally, using SERVAL we recomputed stellar activity index time series 
derived from the available CARMENES spectra: the chromatic index (CRX), 
the differential line width (dLW), calcium infra-red triplet (cairt), and the H$\alpha$ measurements. 
For a detailed description of the SERVAL activity indicators we refer to \citet{Zechmeister2018}.
All CARMENES Doppler measurements, activity index data, and their individual formal uncertainties 
used for our analysis are available in the Appendix (Tables \ref{tab:CARM_VIS}-\ref{tab:CARM_NIR2}).\looseness=-8

\section{Data analysis}
\label{Sec4}
 
\subsection{Periodogram analysis}

\Autoref{s_h_index}, left panels, show generalized Lomb-Scargle \citep[GLS;][]{Zechmeister2009}  periodograms of the HIRES and CARMENES time series  alongside the GLS periodograms of the radial velocity residuals as obtained using a pre-whitening\footnote{Consecutive signal subtraction until no
significant peaks are left in the data residuals \citep{Hatzes2013}. 
For  GJ\,1148,
we subtracted the full Keplerian planetary signals to account for their non-zero eccentricities.} procedure.
The panels of  \autoref{s_h_index} are labeled to indicate the HIRES and CARMENES GLS periodograms.
The residuals to a flat model to the HIRES data (i.e.,\ no planet, only RV offset), 
show a strong power at a period of 41.3 days, which when 
fitted with a planet signal yields an additional residual signal with period of $\sim$ 530\,d.
These are the signals of GJ\,1148 b and c \citep[for details see][]{Trifonov2018a}.

As in the HIRES data, the new CARMENES data are
consistent with the two planetary signals with periods of $\sim$ 41.3\,d and 532.6\,d, respectively. 
There are no significant GLS peaks in the residuals of the simultaneous 
two-planet Keplerian fit applied to the HIRES and CARMENES optical data. 
In \autoref{s_h_index} the adopted stellar rotational period of 71.5\,d from 
\citet{DiezAlonso2019} of GJ\,1148 is shown with a red dashed line in each panel.

The HIRES S- and H-index time series of \citet{Butler2017} lack any significant periodicity that could be associated with the two strong planetary signals seen in the RVs \citep{Trifonov2018a}.
From \autoref{s_h_index}, it is evident that 
the spectral activity indicator time series from 
CARMENES also do not yield significant peaks at the known planetary periods (blue dashed lines).

    \begin{table}[] % !!!
    % \begin{adjustwidth}{-4.0cm}{} 
    % \resizebox{0.69\textheight}{!}
    % {\begin{minipage}{1.1\textwidth}
    \centering   
    \caption{{Dynamical best-fit osculating parameters of the GJ\,1148 b and c planets for a coplanar edge-on configuration. Orbital elements are in the Jacobi frame and are valid for the first HIRES observational epoch BJD =  2451581.046. The dynamical modeling of the combined HIRES and CARMENES data was unable to constrain the mutual inclination.}}   
    \label{table:2}      
    \begin{tabular}{lrrrrrrrr}     % 2 columns 
    \hline
    \hline  
    \noalign{\vskip 0.7mm}      
    Parameter \hspace{0.0 mm}&\hspace{6.0 mm} Planet b & Planet c \\
    \hline 
    \noalign{\vskip 0.7mm} 
        $K$  [m\,s$^{-1}$]            &     38.54$_{-0.37}^{+0.43}$ &     12.26$_{-0.56}^{+0.59}$ \\ \noalign{\vskip 0.9mm}
        $P$  [day]                    &     41.380$_{-0.002}^{+0.001}$ &    532.635$_{-1.094}^{+0.994}$ \\ \noalign{\vskip 0.9mm}
        $e$                           &      0.375$_{-0.009}^{+0.008}$ &      0.375$_{-0.041}^{+0.036}$ \\ \noalign{\vskip 0.9mm}
        $\varpi$  [deg]               &    258.7$_{-1.6}^{+1.7}$ &    206.6$_{-5.7}^{+6.3}$ \\ \noalign{\vskip 0.9mm}
        $M_{\rm 0}$  [deg]            &    297.8$_{-2.2}^{+2.0}$ &    276.0$_{-8.0}^{+6.9}$ \\ \noalign{\vskip 0.9mm}
        $i$  [deg]                    &     90.0 (fixed)  &     90.0 (fixed) \\ \noalign{\vskip 0.9mm} 
        $\Omega$  [deg]               &      0.0  (fixed) &      0.0 (fixed) \\ \noalign{\vskip 0.9mm} 
        $a$  [au]                     &      0.166$_{-0.001}^{+0.001}$ &      0.910$_{-0.001}^{+0.001}$ \\ \noalign{\vskip 0.9mm} 
        $m \sin i$  [$M_{\rm jup}$]   &      0.304$_{-0.004}^{+0.003}$ &      0.227$_{-0.022}^{+0.001}$ \\ \noalign{\vskip 0.9mm} 
        N$_{\rm RV}$ data             &        \makebox[\dimexpr(\width-3.5em)][l]{201} \\ 
        Systemic velocity $\gamma_{\rm HIRES}$          &      \makebox[\dimexpr(\width-3.5em)][l]{1.96$_{-0.42}^{+0.46}$} \\ \noalign{\vskip 0.9mm}
        Systemic velocity $\gamma_{\rm CARMENES}$       &     \makebox[\dimexpr(\width-3.5em)][l]{-3.40$_{-0.41}^{+0.42}$} \\ \noalign{\vskip 0.9mm}
        RV jitter $\sigma_{\rm HIRES}$          &      \makebox[\dimexpr(\width-3.5em)][l]{3.61$_{-0.20}^{+0.59}$} \\ \noalign{\vskip 0.9mm}
        RV jitter $\sigma_{\rm CARMENES}$       &      \makebox[\dimexpr(\width-3.5em)][l]{1.940$_{-0.07}^{+0.49}$} \\ \noalign{\vskip 0.9mm}
        $rms$ [m\,s$^{-1}$]        &      \makebox[\dimexpr(\width-3.5em)][l]{4.14} \\ \noalign{\vskip 0.9mm}
        $-\ln\mathcal{L}$             &   \makebox[\dimexpr(\width-3.5em)][l]{-549.75} \\
        $\Delta\mathcal{L}$             &   \makebox[\dimexpr(\width-3.5em)][l]{379.91} \\
        
\hline 
\noalign{\vskip 0.7mm} 
\end{tabular}  
% \end{minipage}}
% \end{adjustwidth}
%\tablefoot{\small }
\end{table}

\begin{figure*}
\sidecaption
\includegraphics[width=12cm]{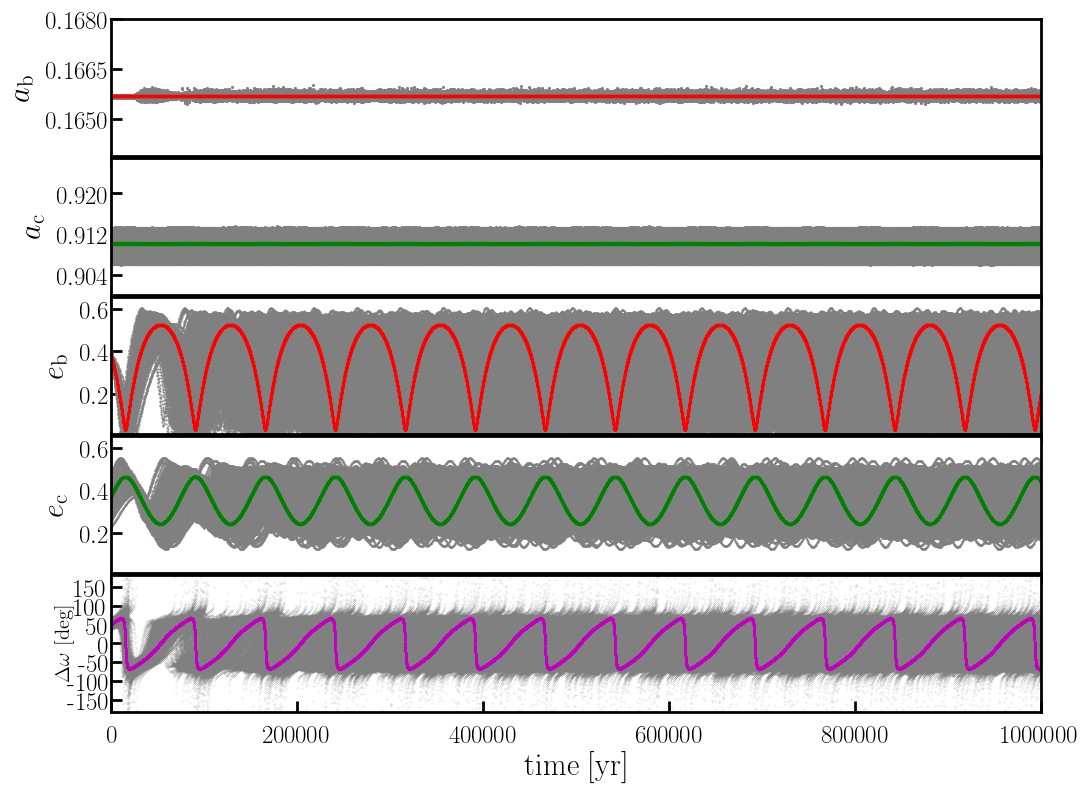} 
     \caption{Orbital evolution of the GJ\,1148 system for a Myr long N-body integration using Wisdom-Holman scheme.
     From top to bottom: Evolution of the planetary semi-major axes, eccentricities,
     and the apsidal alignment angle $\Delta\omega$ = $\varpi_{\rm b} - \varpi_{\rm c}$.
     The gray area is composed of 1000 randomly chosen stable configurations from the MCMC 
     run, which represent the possible orbital outcome consistent with the RV data. 
     The solid color curves show the evolution of the best-fit configuration. 
     }
   \label{orb_evol}
\end{figure*}

Photometric observations of GJ\,1148 were performed by \citet{Haghighipour2010}, who found a periodic signal at $P_{\rm rot}$ = 98.1\,d, suggesting that this is the likely rotational period of the star (see Figure~5 and Section~5 in their paper for details).
However, from the Hungarian-made Automated Telescope Network \citep[HATNet;][]{Bakos2004} survey of field K and M dwarfs, \citet{Hartman2011} estimated a rotational period for GJ\,1148 of $P_{\rm rot}$ = 73.5\,d. 
This value agrees within the uncertainty reported by \citet{DiezAlonso2019}, $P_{\rm rot}$ = 71.5$\pm$5.1\,d, who performed more detailed 
analysis of the SuperWASP \citep{Pollacco2006}, the Northern Sky Variability Survey \citep[NSVS;][]{Wozniak2004}, and the HATNet photometry of GJ\,1148.

In the three bottom right panels of \autoref{s_h_index}, we show the GLS periodograms of the SuperWASP, the NSVS light-curve data of GJ\,1148 compiled by \citet{DiezAlonso2019}, and the HATNet photometry.
These photometric measurements were taken with a higher cadence when compared to the RV data, and thus many frequencies in the GLS power spectrum appear significant.
A well-defined significant peak is only detected in the HATNet time series,
whose maximum power is within the $P_{\rm rot}$ = 71.5$\pm$5.1\,d uncertainties 
(red shaded frequency range in \autoref{s_h_index}) estimated by \citet{DiezAlonso2019} based on the three photometry data sets.
However, there is no definitive agreement between the SuperWASP, NSVS, and the HATNet photometry on the one hand, and the spectral Doppler measurements and activity indicators on the other. 
We concluded that there is no photometric periodic signal that could be firmly associated with the GJ\,1148~b and~c planets.

\begin{table}
    \centering   
    \caption{{Statistical properties of the CARMENES near-IR data tested against a flat model (null), flat model with jitter, static one-planet model assuming only GJ\,1148 b,
    and the two-planet best-fit model adopted from the optical RV analysis. The only free parameter in the null model is the near-IR RV offset, while  in the rest  RV offset and the jitter are free.
    }
    }  
    \label{table:nir}      
    \begin{tabular}{@{}lcccc@{}}
    \hline
    \hline  
    \noalign{\vskip 0.7mm}      
    Stat.  & null  & Jitt. only \hspace{0.0 mm}        & GJ\,1148 b & GJ\,1148 b \& c\\
     \noalign{\vskip 0.7mm} 
   \hline 
    \noalign{\vskip 0.7mm} 
        $\chi^2$                 &        950.28  & 68.08 &    72.43 &    65.74    \\\noalign{\vskip 0.7mm} 
 %       $\chi_{\nu}^2$                &      1.126 &    210.529 &    210.529\\
        $rms$ [m\,s$^{-1}$]   &        25.36  &  25.30 &    9.34 &    8.06      \\\noalign{\vskip 0.7mm} 
        $-\ln\mathcal{L}$         &       671.15  & 315.94&    239.86 &    228.88  \\\noalign{\vskip 0.7mm} 
        $\Delta\mathcal{L}$         &       \dots  & 355.21 &   431.29 &    442.27 \\\noalign{\vskip 0.7mm} 
         p-value$^a$              &        \dots & 1.7$^{-39}$ &  1.3$^{-38}$ &  5.3$^{-40}$   \\\noalign{\vskip 0.7mm}        

        $\gamma$ [m\,s$^{-1}$]           &         $-$4.98 & $-$3.36 &    $-$3.33 &    $-$7.63     \\\noalign{\vskip 0.7mm} 
         $\sigma$ [m\,s$^{-1}$]           &         0 & 23.84 &   3.27 &    0.01    \\\noalign{\vskip 0.7mm}         
        \hline %\noalign{\vskip 0.7mm} 
    \end{tabular}  
    \tablefoot{\small $a$ -- calculated via F-test against the Null model.}
\end{table}

\subsection{Orbital update of the GJ\,1148 system}
\label{Sec4.1}

For the analysis of the precise Doppler data of GJ\,1148 we use {\em The Exo-Striker}
fitting toolbox\footnote{{\em The Exo-Striker} is freely 
available at \url{https://github.com/3fon3fonov/exostriker}} \citep{Trifonov2019_es}.
{\em The Exo-Striker} provides a large variety of fitting and sampling 
schemes coupled either with a standard Keplerian model or with a more complex 
N-body model. The latter takes into account the gravitational perturbations 
that occur in multi-planet systems while it models the radial component of the stellar velocity
caused by the companions. 
For a widely separated system such as GJ\,1148 there is little or no practical
incentive in using the more expensive dynamical model over the simple Keplerian model.
The  Keplerian and N-body models are indistinguishable within the temporal baseline of the combined HIRES and CARMENES observations \citep{Trifonov2018a}.  
The dynamical model, however, has the advantage of being able to fit for mutually inclined orbits 
which distribution and long-period stability we aim to study.
Therefore, our natural choice is to use the N-body RV model for the orbital analysis of the GJ\,1148 system.%\looseness=-19

To derive the best fit we adopt a maximum likelihood estimator (MLE) scheme, 
which optimizes the N-body model's likelihood value ($-\ln \mathcal{L}$)   via the Nelder-Mead algorithm \citep[also known as Simplex,][]{NelderMead}.
As in \citet{Trifonov2018a} we optimize the osculating semi-amplitude $K$, orbital period $P$, 
eccentricity $e$, argument of periastron $\omega$, and mean anomaly $M$ 
for the first HIRES observational epoch and the HIRES and CARMENES data velocity offsets.
In these analyses we also include the HIRES and CARMENES velocity jitter
variance as additional fitting parameters \citep{Baluev2009}. 
Additionally, when we study mutually inclined orbits we also fit for the 
orbital inclinations $i$ and the difference of the orbital lines of node $\Delta\Omega$.
To estimate the parameter uncertainties of our best fits and to perform a  parameter distribution analysis, we rely on a Markov chain Monte Carlo 
(MCMC) sampling using the $emcee$ sampler \citep{emcee}, 
which is integrated within {\em The Exo-Striker}.

Our new best-fit analysis is largely consistent with our previous results for the GJ\,1148 system presented in \citet{Trifonov2018a}.
Assuming an edge-on coplanar configuration, our new N-body model fitted to the CARMENES and HIRES data yields
osculating orbital parameters for the inner planet GJ\,1148 b:
$K_{\rm b}$ = 38.54$_{-0.37}^{+0.43}$  m\,s$^{-1}$,
$P_{\rm b}$ = 41.380$_{-0.002}^{+0.001}$\,d,
$e_{\rm b}$ = 0.375$_{-0.009}^{+0.008}$,  
and for GJ\,1148~c:  
$K_{\rm c}$ = 12.26$_{-0.56}^{+0.59}$ m\,s$^{-1}$,
$P_{\rm c}$ = 532.6$_{-1.1}^{+1.1}$\,d,
$e_{\rm c}$ = 0.375$_{-0.041}^{+0.036}$. 
From the updated stellar mass of GJ\,1148 and these best-fit estimates we derive minimum planetary masses of  
$m_b \sin i$  = 0.304 $M_{\rm Jup}$ and $m_c \sin i$ = 0.227 $M_{\rm Jup}$, % (96.7 and 68.1 $M_\oplus$),
and semi-major axes $a_{\rm b}$ = 0.166 au and $a_{\rm b}$ = 0.910 au, respectively.
\Autoref{table:2} summarizes the best-fit parameters and uncertainties of our new coplanar edge-on N-body model of the GJ\,1148 system. 
The best fit to the new CARMENES and HIRES RV time series is shown 
in \autoref{new_data}, together with a phase-folded representation of the 
GJ\,1148 b and~c planetary signals. From \autoref{new_data} it is clear that 
the CARMENES and HIRES data are fully consistent with the presence of the two planetary signals.\looseness=-8

\subsection{CARMENES near-IR data}
\label{sec:CARM_NIR}

We tested the 68 CARMENES near-IR RVs for consistency with the two-planet signals. 
The near-IR data was not used in our analysis together with the available optical data, for two reasons.
First,   the precision of the CARMENES near-IR RV data of GJ\,1148 is significantly lower than that achieved in the optical channel of CARMENES.
This is expected for GJ\,1148:  for stars of spectral class M\,4.0\,V the
spectroscopic information needed for precise RV measurements is still more 
abundant in the CARMENES-VIS spectra \citep[see Fig.~6 in][]{Reiners2018b}.
Thus, the weight of the near-IR data is much lower in the modeling.
Second,   the CARMENES optical and NIR data are taken practically simultaneously (usually with a very small deviation 
in their time stamps due to the slightly different photon mid-point center). 
This causes difficulties for the N-body modeling, which must automatically adapt the integration 
time steps to very small values in order to go between two neighboring CARMENES data points.
This makes the modeling of the near-IR data together with the higher quality optical data impractical. 
Instead, we decided to test the  near-IR data  against the best-fit model constructed from the optical, 
following the methodology adopted by \citet{Trifonov2015} for the VLT-CRIRES \citep{Kaeufl} near-IR data.
So we applied this model to the near-IR data by optimizing only the RV offset and we studied its statistical properties.

\Autoref{NIR_data} shows the CARMENES near-IR RV measurements and a phase-folded 
representation of the GJ\,1148 b and~c planetary signals from the best-fit model from \Autoref{table:2}.
A visual inspection shows that the signal of GJ\,1148 b is clearly present in the CARMENES-NIR data. 
The GJ\,1148 c planetary signal is somewhat sparsely sampled around  periastron by the CARMENES-NIR data, 
yet there is  good agreement in phase and amplitude between the data and the model.

\begin{figure}
    \includegraphics[width=\linewidth]{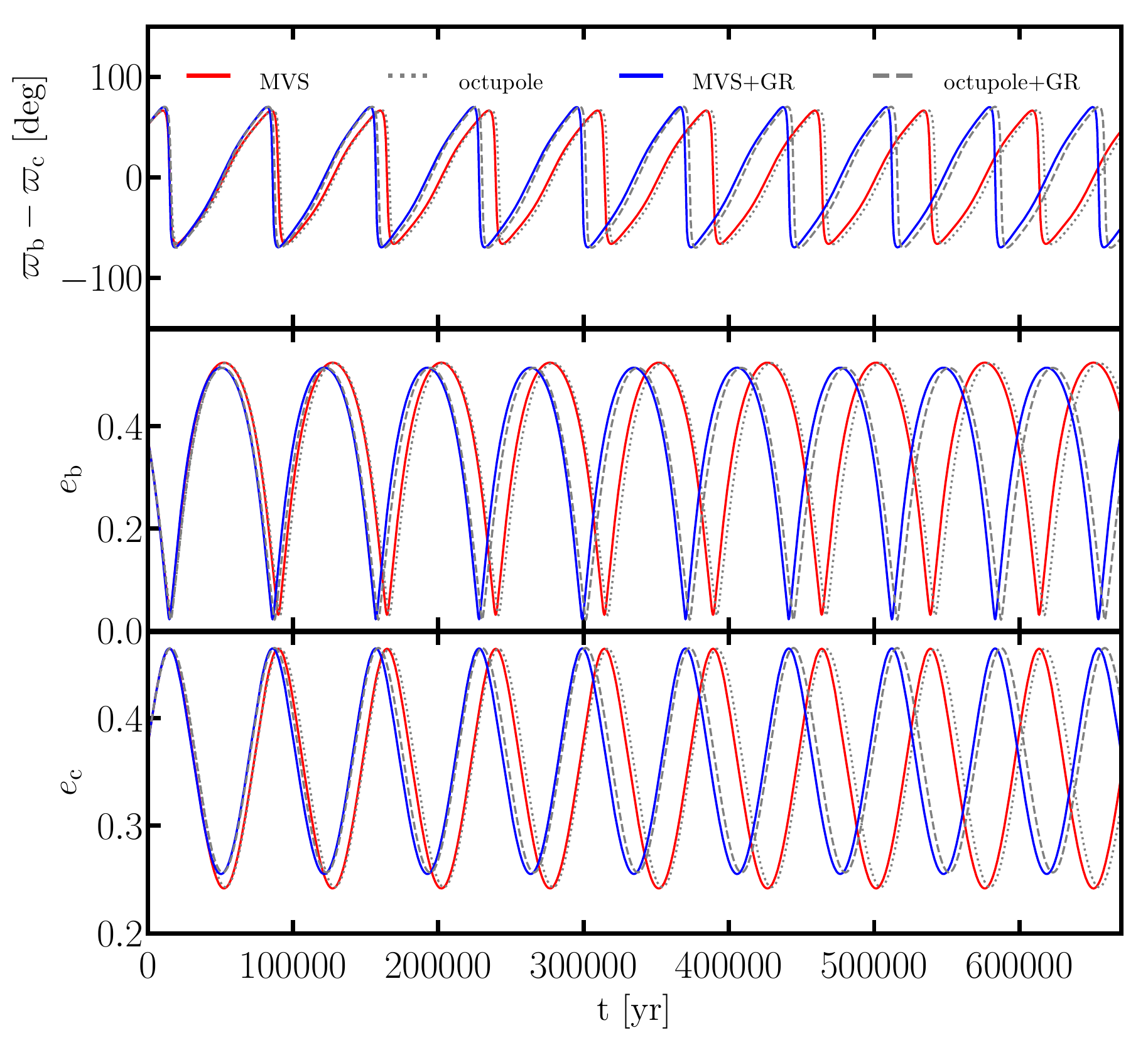} 
    \caption{
    Evolution of the orbital eccentricities $e_{\rm b}$ and $e_{\rm c}$ and 
    apsidal alignment angle $\Delta\varpi$ = $\varpi_{\rm b}$ - $\varpi_{\rm c}$.
Red curves represent the evolution obtained from the Wisdom-Holman N-body integration 
evolution of the best fit (see \autoref{orb_evol}), while red curves
are the  Wisdom-Holman N-body integration with GR precession included.
 Gray dashed and dotted curves represent the evolution of the best fit but derived from the octupole secular perturbation theory and the octupole 
evolution with a GR precession correction, respectively.\looseness=-5
    }
    \label{oct_evol}
\end{figure}

The statistical properties of the near-IR data with respect to the planetary signals of GJ\,1148 are summarized in \autoref{table:nir}. 
Assuming no planets (i.e.,\ the null hypothesis), the near-IR data have a large $rms$ = 25.36 m\,s$^{-1}$,
while subtracting only the innermost planetary signal from the data is strongly preferred yielding an
$rms$ = 9.34 m\,s$^{-1}$, and a $p$ = 1.3$^{-38}$. 
Finally, applying the best-fit two-planet model results in an $rms$ = 8.06 m\,s$^{-1}$ and $p$ = 5.3$^{-40}$ 
with respect to the flat model.
 We note that in the two-planet model the CARMENES-NIR jitter is 
only 0.01 m\,s$^{-1}$, meaning that the internal data uncertainties ($\hat\sigma_{\rm NIR}$ = 7.7  m\,s$^{-1}$)
are adequately estimated.
In terms of relative probability, $R$ = $e^{\Delta\ln\mathcal{L}}$, between a model that assumes only the 
presence of GJ\,1148 b and the best-fit two-planet model, requires $\Delta\ln\mathcal{L} > 5$ to claim significance \citep[see][]{Baluev2009}
The best two-planet model applied to the near-IR data yields $\Delta\ln\mathcal{L}$ $\sim$ 11 (see \autoref{table:2}) over the 
one-planet model,  making the two-planet model the much better choice.
We therefore conclude that the CARMENES near-IR RVs fully agree with the best-fit model 
from the CARMENES optical data and the HIRES data. 
Therefore, the near-IR data strengthens the multi-planet hypothesis around GJ\,1148.

\section{Dynamics}
\label{Sec5} 

\subsection{Coplanar edge-on configurations}
\label{Sec5.1}

In \citet{Trifonov2018a} we  show that the  best-fit 
orbital configuration of the GJ 1148 system is stable for at least 10\,Myr, but exhibits 
strong long-term secular dynamical 
interactions leading to large oscillations of the planetary eccentricities on a 
 timescale of $\sim$80\,000 yr. 
In our new dynamical analysis, we inspect  the dynamics of the new best fit, and also 
the overall stability of the MCMC posterior parameter distribution consistent with the data.

Dynamical analyses are performed using a custom version of the 
Wisdom-Holman algorithm \citep[][]{Wisdom1991}, 
which works in the Jacobi coordinate system \citep[e.g.,][]{Lee2003}. 
Our long-term stability integrations are performed for a maximum 
of 10\,Myr with an integration time step equal to 0.4\,d.
We find this time step and maximum integration time to be sufficient
for an accurate dynamical test of the system.  
For each MCMC integration, we automatically monitor the evolution of the planetary 
semi-major axes and eccentricities as a function of time to ensure that the system remains
regular and well separated at any given time of the simulation.
Any deviation of the planetary semi-major axes by more than 20\% from
their starting values, or eccentricities leading to crossing orbits, were considered unstable.\looseness=-2

The coplanar edge-on MCMC posterior distributions around the best fit from \autoref{table:2}, 
and their respective correlations are shown in \autoref{FigApp_mcmc}. 
We found that all the coplanar edge-on MCMC samples are stable for 10\,Myr. 
This is perhaps not a surprise given the relatively large orbital separation between the two massive planets
and taking into account that the best-fit model is very well constrained by the RV data.
From the coplanar edge-on perspective, we cannot draw conclusions based upon long-term stability, 
but the overall dynamics of the MCMC posterior parameter distribution allow us to 
to shed more light on the possible dynamical architecture of the system.

\Autoref{orb_evol} shows the 1\,Myr orbital evolution of 
the best fit and 1000 randomly chosen configurations from our MCMC test.
The planetary semi-major axes remain nearly constant with no major 
deviations from the starting values over the integration.
The eccentricities, however, exhibit large variations in the range of $e_{\rm b}$ = 0.01 to 0.60 and $e_{\rm c}$ = 0.10 to 0.50.
Interestingly, most of the MCMC configurations exhibit apsidal alignment with the 
angle $\Delta\omega$ = $\varpi_{\rm b} - \varpi_{\rm c}$ librating around 0$^\circ$ with a typical semi-amplitude of 60$^\circ$;
  very few configurations circulate between 0$^\circ$ and 360$^\circ$.
This dynamical behavior can be nicely explained using the octupole secular theory.
For comparison, we also integrate the time evolution of the orbital elements for the planar hierarchical three-body problem, using the 
octupole-level secular perturbation equations derived by \citet{Ford2000}.

\begin{figure}
    \includegraphics[width=9cm]{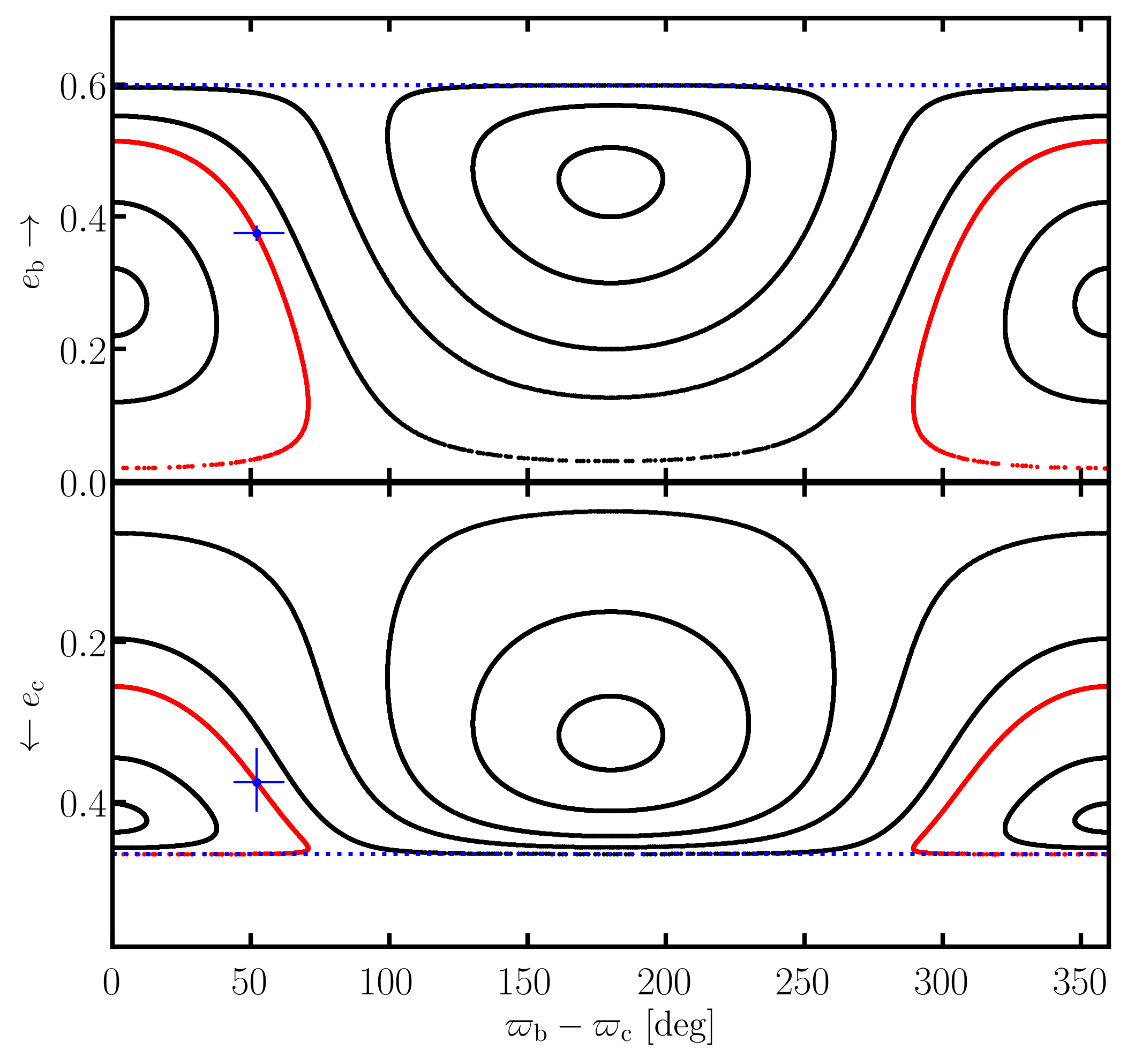} 
    \caption{
    Trajectories in the phase-space diagrams of the planetary eccentricities 
$e_{\rm b}$ and $e_{\rm c}$ vs. the 
apsidal alignment angle $\Delta\varpi$ = $\varpi_{\rm b}$ - $\varpi_{\rm c}$  
for the GJ\,1148 system constructed from the octupole-level 
secular perturbation theory, with GR correction included, assuming the same total angular momentum.
Blue crosses indicate the N-body best-fit parameters of $e_{\rm b}$, $e_{\rm c}$ 
and $\varpi_{\rm b}$ - $\varpi_{\rm c}$ and their uncertainties,
while the red paths represent the trajectories starting at the best fit.
% For the other trajectories the initial conditions are \dots
The blue dotted lines represent the maximum planetary eccentricities, which can be attained
for the total angular momentum of the system.
    }
    \label{oct_evol_traj}
\end{figure}

\begin{figure*}
    \sidecaption
    \includegraphics[width=12cm]{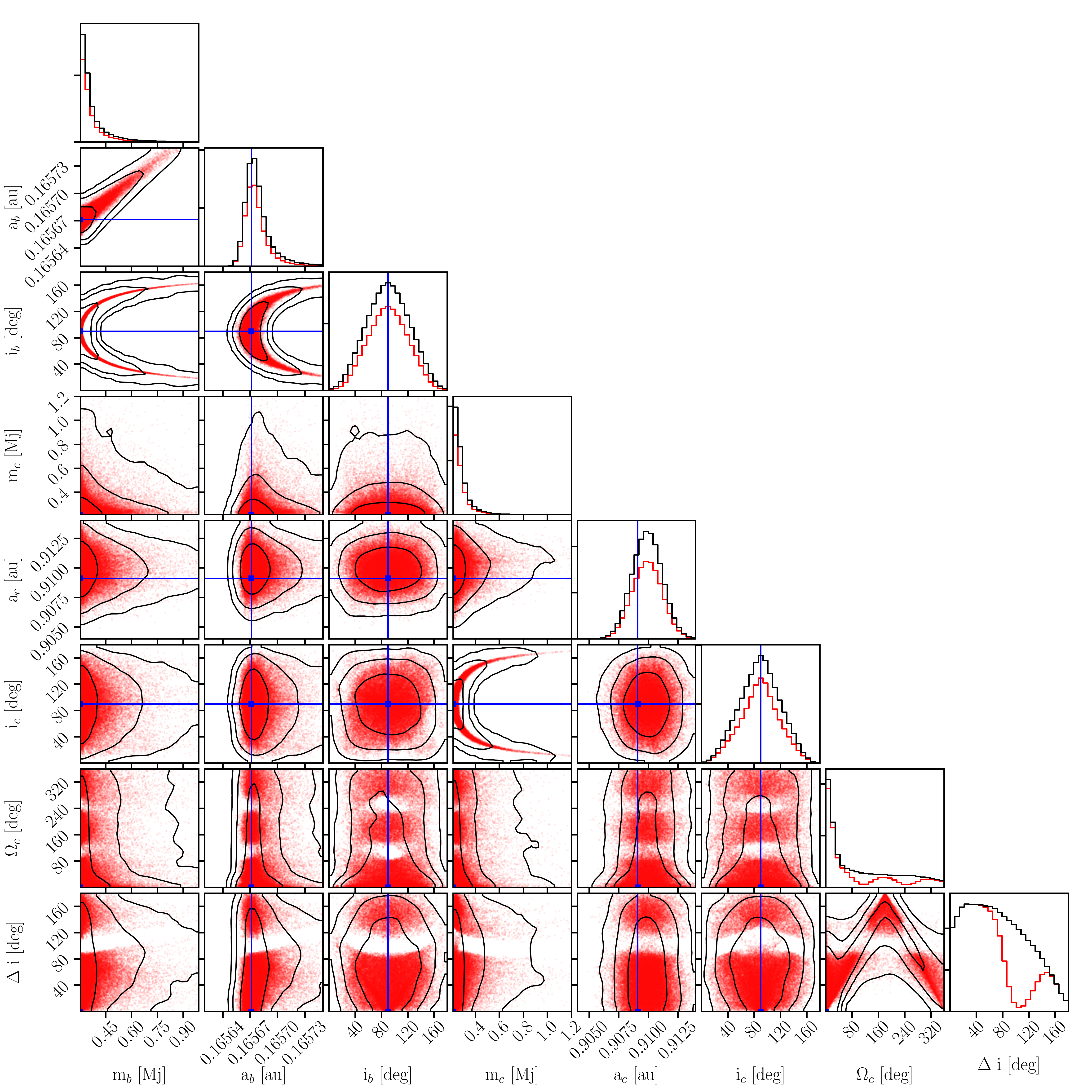} 
    \caption{Dynamical MCMC samples of the GJ\,1148 two-planet system allowing for mutual inclinations. 
    Red samples represent the configurations which survived 10 Myr without disrupting the system.
    The MCMC samples peak at $\Delta i \sim$ 30$^\circ$, where most of the stable solutions are found. A second smaller peak of stable solutions appears at $\Delta i \sim$ 150$^\circ$ (i.e., retrograde configuration).
    Near polar configurations are dynamically unstable due to strong Lidov-Kozai effects.
    }
    \label{mcmc_mut_incl}
\end{figure*}

In \autoref{oct_evol} we show the evolution of the planetary eccentricities $e_{\rm b}$ and $e_{\rm c}$ 
and the apsidal alignment angle $\Delta\varpi$ = $\varpi_{\rm b}$ - $\varpi_{\rm c}$. We integrate the best-fit solution both with N-body and secular codes. 
For the secular code, we use \texttt{SecuLab}\footnote{\href{https://github.com/eugeneg88/SecuLab}{https://github.com/eugeneg88/SecuLab}} up to octupole order, with corrected double averaging terms \citep{luo16, grishin18}. Another octopule code without the corrected terms \citep{Lee2003} gives almost identical results. The linear secular theory (up to second order in eccentricities) predicts a secular period of $73400$\,yr (see Eq.~7.9-7.28 in \citealp{md99}). The secular integration yields a timescale of $75800$\,yr. The differences are due to the large eccentricities of both planets. 

General relativistic precession can also play a role in the long-term evolution of the system. The apsidal precession rate of planet b is \citep{Blaes2002,lml15} 
\begin{equation}
    \begin{split}
    \dot{\omega}_{\rm GR} = \frac{6\pi}{P_b} \frac{G M_{\star}}{c^2 a_b (1-e_b^2)} 
    \end{split}
    \label{eq:omegaGR}
\end{equation}
and the typical timescale is $T_{\rm GR} = 2\pi/\dot{\omega}_{\rm GR} \approx 1.54$ Myr. Including general relativity (GR) precession in the secular code yields a slightly shorter secular period of $72500$\,yr. The simplified expression for the secular timescale with GR is $T_{\rm sec,GR}=T_{\rm sec}/(1+T_{\rm sec}/T_{\rm GR})\approx 72200$\,yr, which slightly underestimates the timescale obtained from numerical integration. This is due to the large variations in the eccentricity, which affects the GR precession rate as $\propto (1-e_b^2)^{-1}$. 

Overall, the Wisdom-Holman N-body code and the secular code are consistent with each other.
Apart from the slightly different timescales, the phase and amplitude of the osculating parameters of the octupole prediction are clearly consistent with those from the N-body dynamical evolution of the GJ\,1148 system, including GR corrections. Moreover, the 
Wisdom-Holman code with GR corrections applied on the GJ\,1148 system  is the first public code available which has been compared to secular theory. Only very recently have other public N-body codes  started incorporating GR corrections \citep{Tamayo2019}.

\Autoref{oct_evol_traj} shows the dynamical behavior of the planetary eccentricities and apsidal alignment for the co-planar 
configuration in the phase-space diagrams of $e_{\rm b}$ and $e_{\rm b}$  on the y-axis and the
apsidal alignment angle $\Delta\varpi$ = $\varpi_{\rm b}$ - $\varpi_{\rm c}$ on the x-axis. We show the trajectories calculated from the octupole-level 
secular perturbation theory with GR precession, assuming the same total angular momentum as for the GJ\,1148 system,
but with different initial $e_{\rm b}$, $e_{\rm c}$, and $\Delta\varpi$.
The blue crosses in \autoref{oct_evol_traj} indicate the position of the best fit and its uncertainties,
while the red trajectory is the evolution starting from the best fit.
\autoref{oct_evol_traj} shows that $\Delta\varpi$ of the GJ\,1148 system is librating about $0^\circ$ with almost the largest possible amplitude, without going into circulation.

\begin{figure*}[tp]
    \centering
$
\begin{array}{cc} 
    \includegraphics[width=5cm]{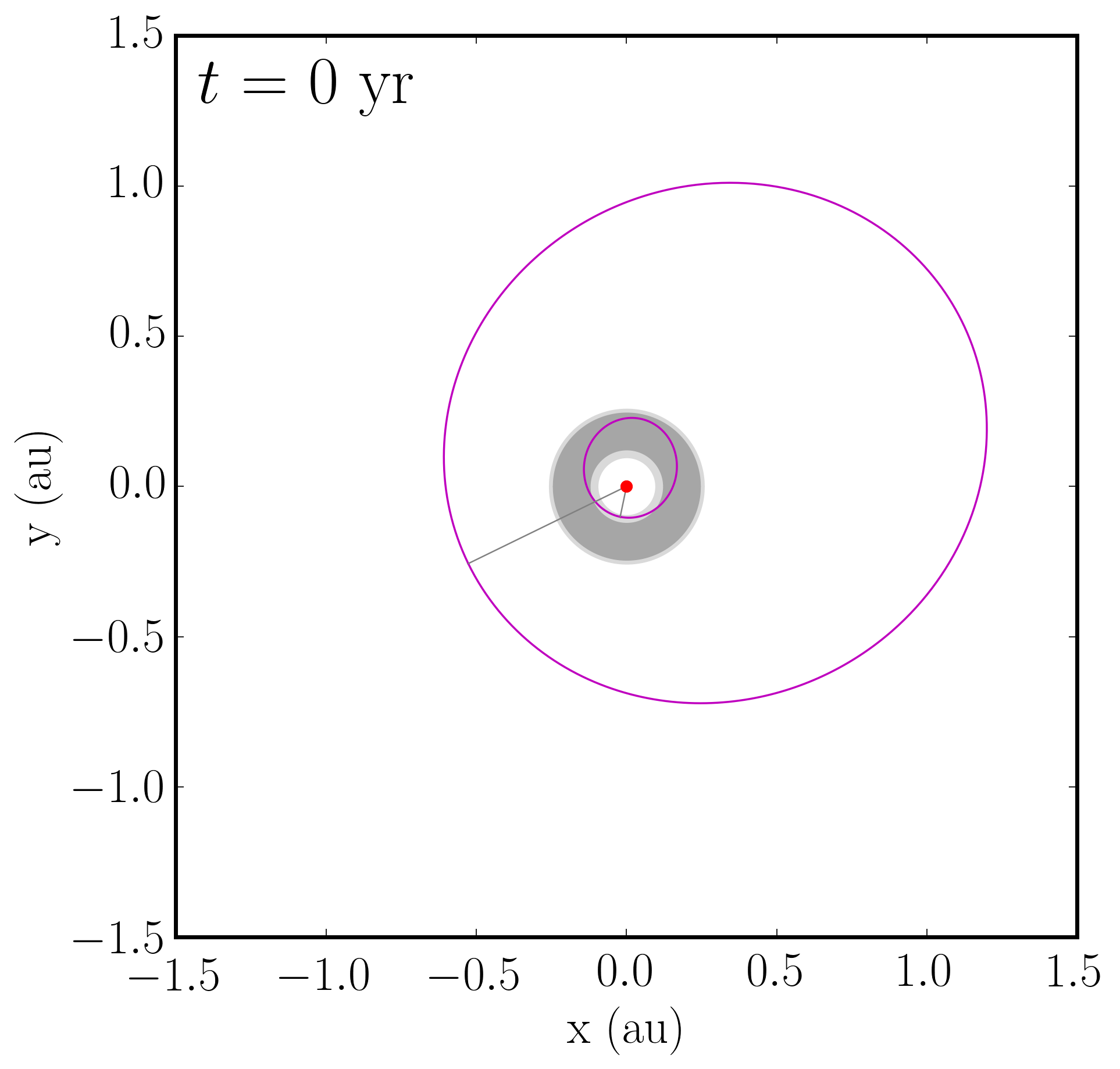} 
    \includegraphics[width=5cm]{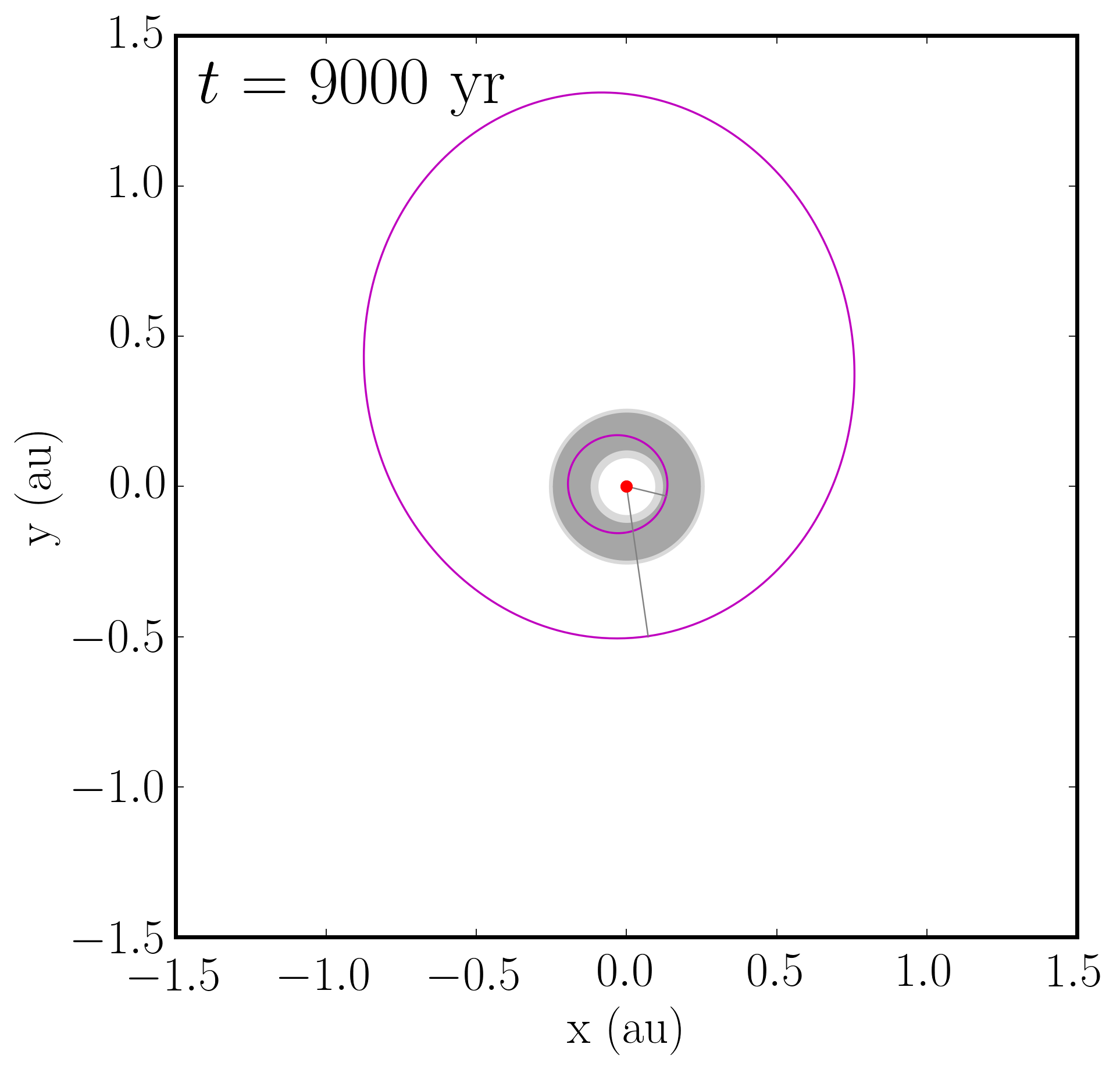}  
    \includegraphics[width=5cm]{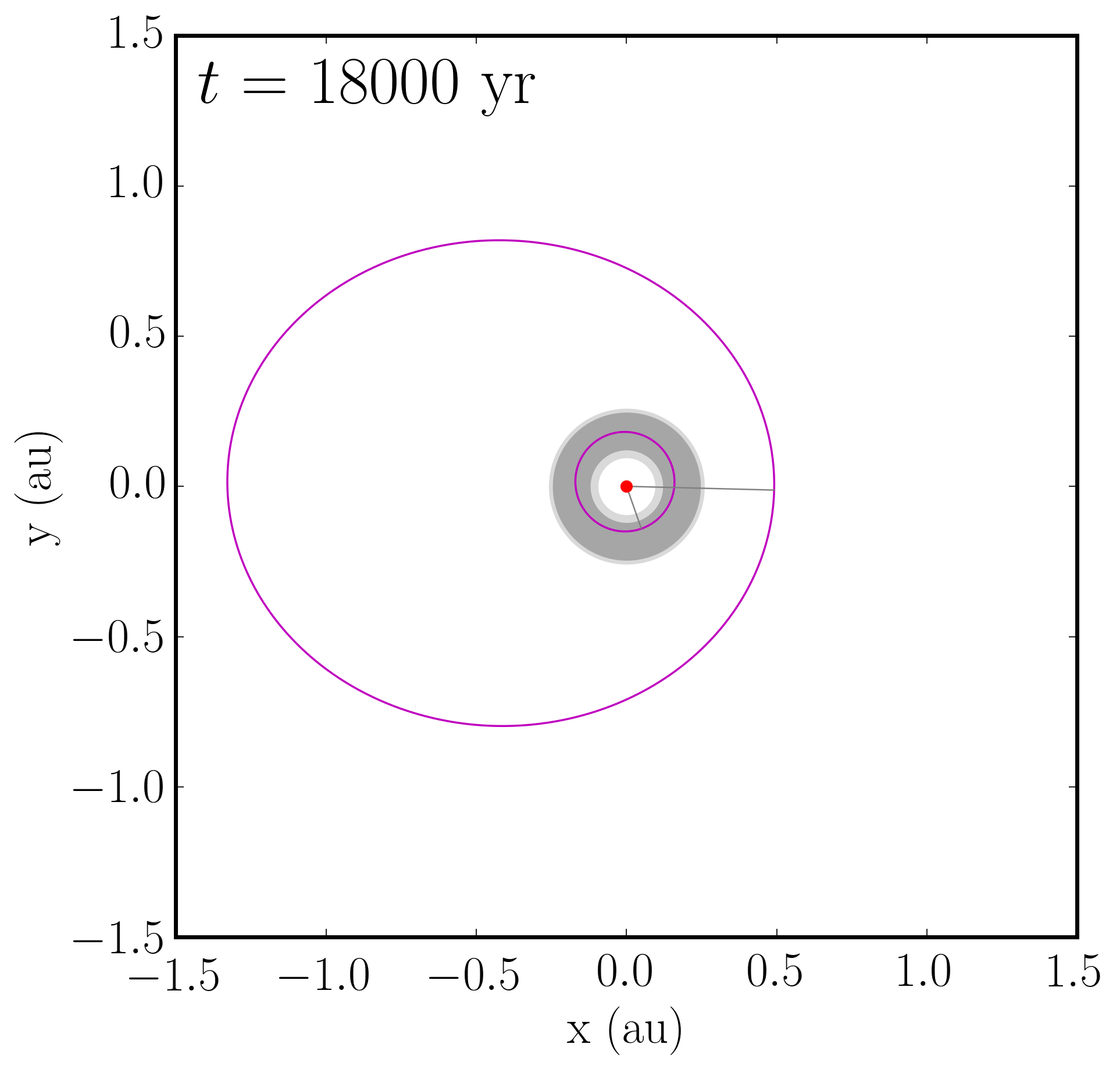} \\
    \includegraphics[width=5cm]{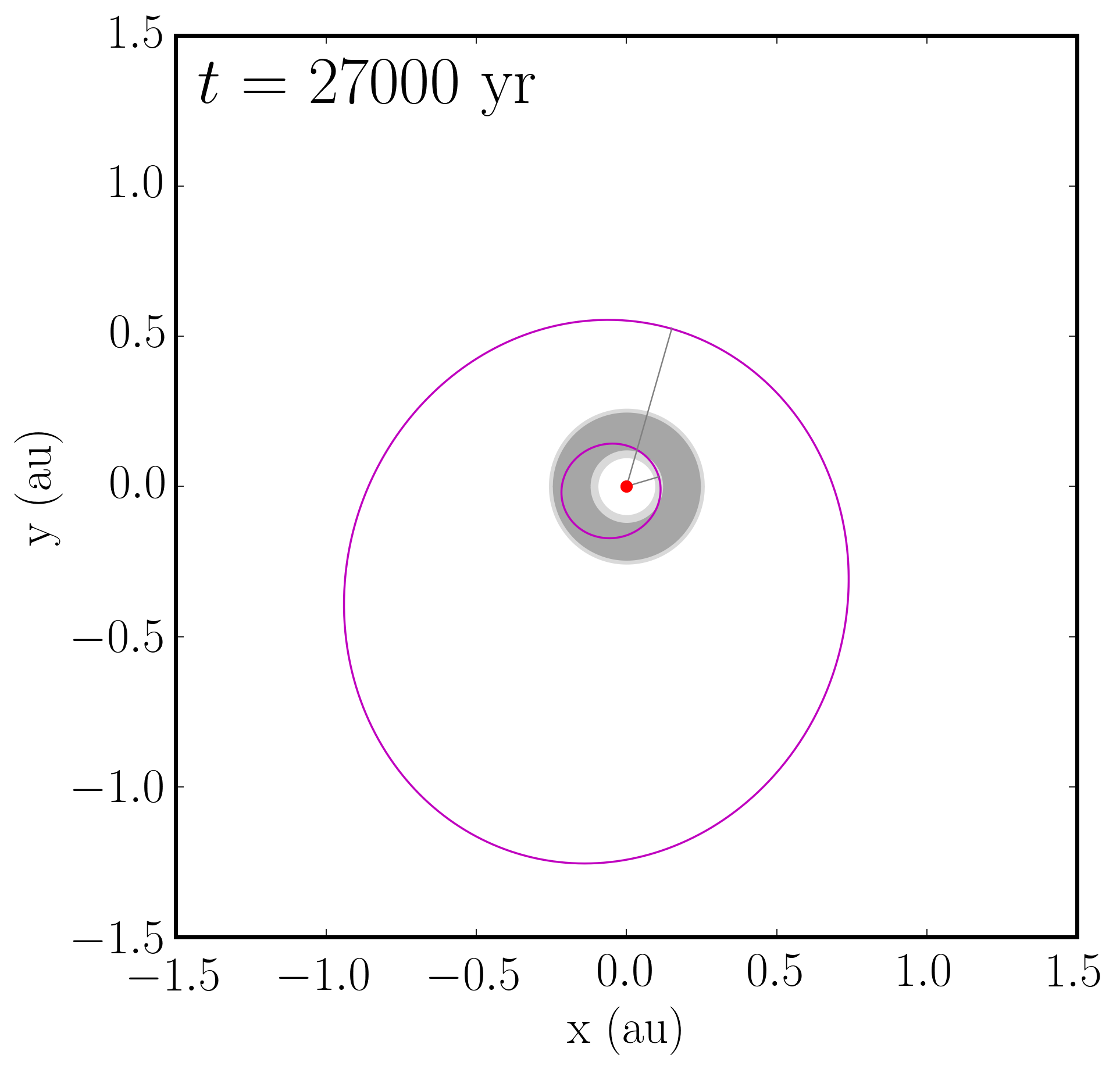} 
    \includegraphics[width=5cm]{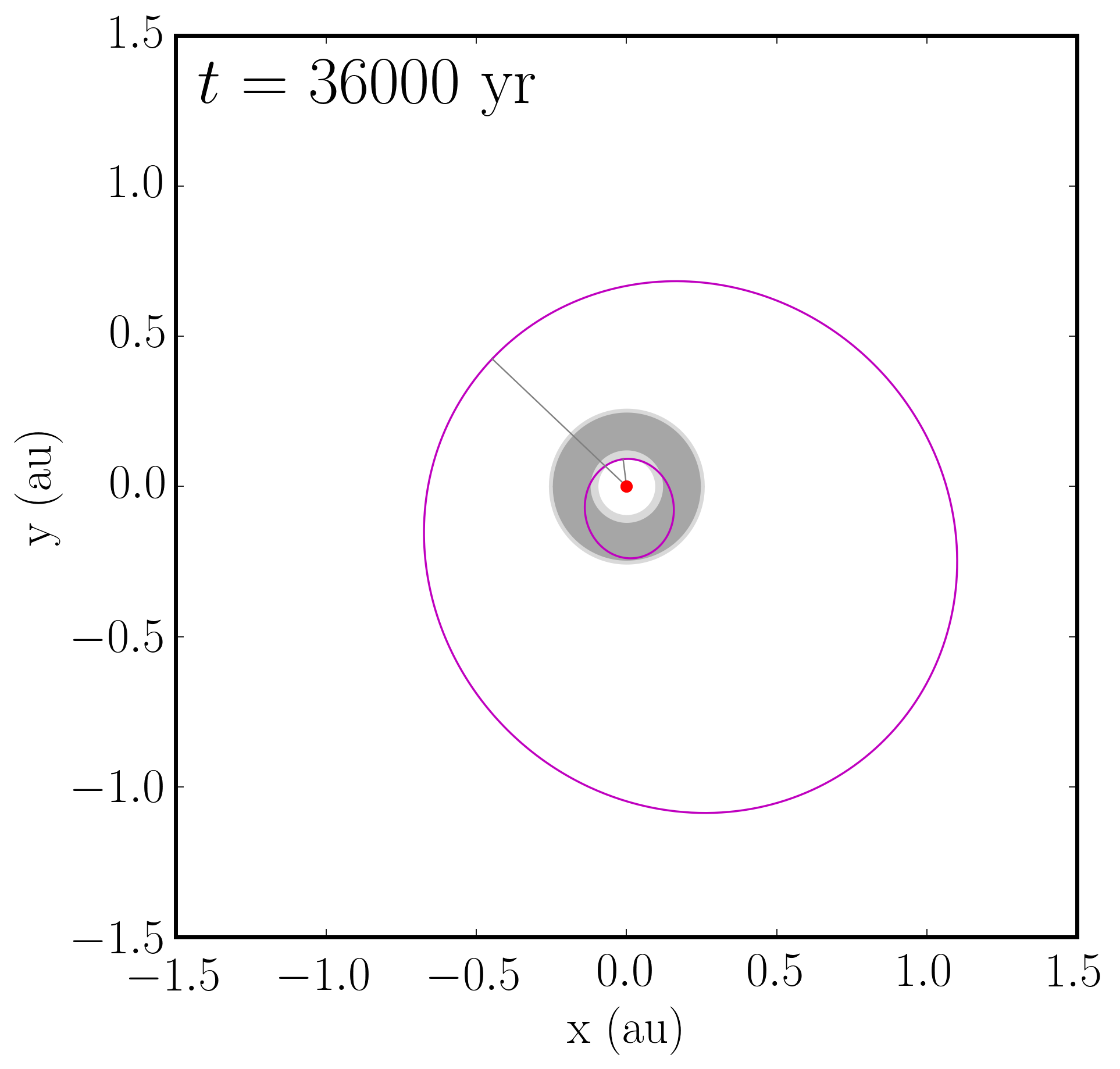} 
    \includegraphics[width=5cm]{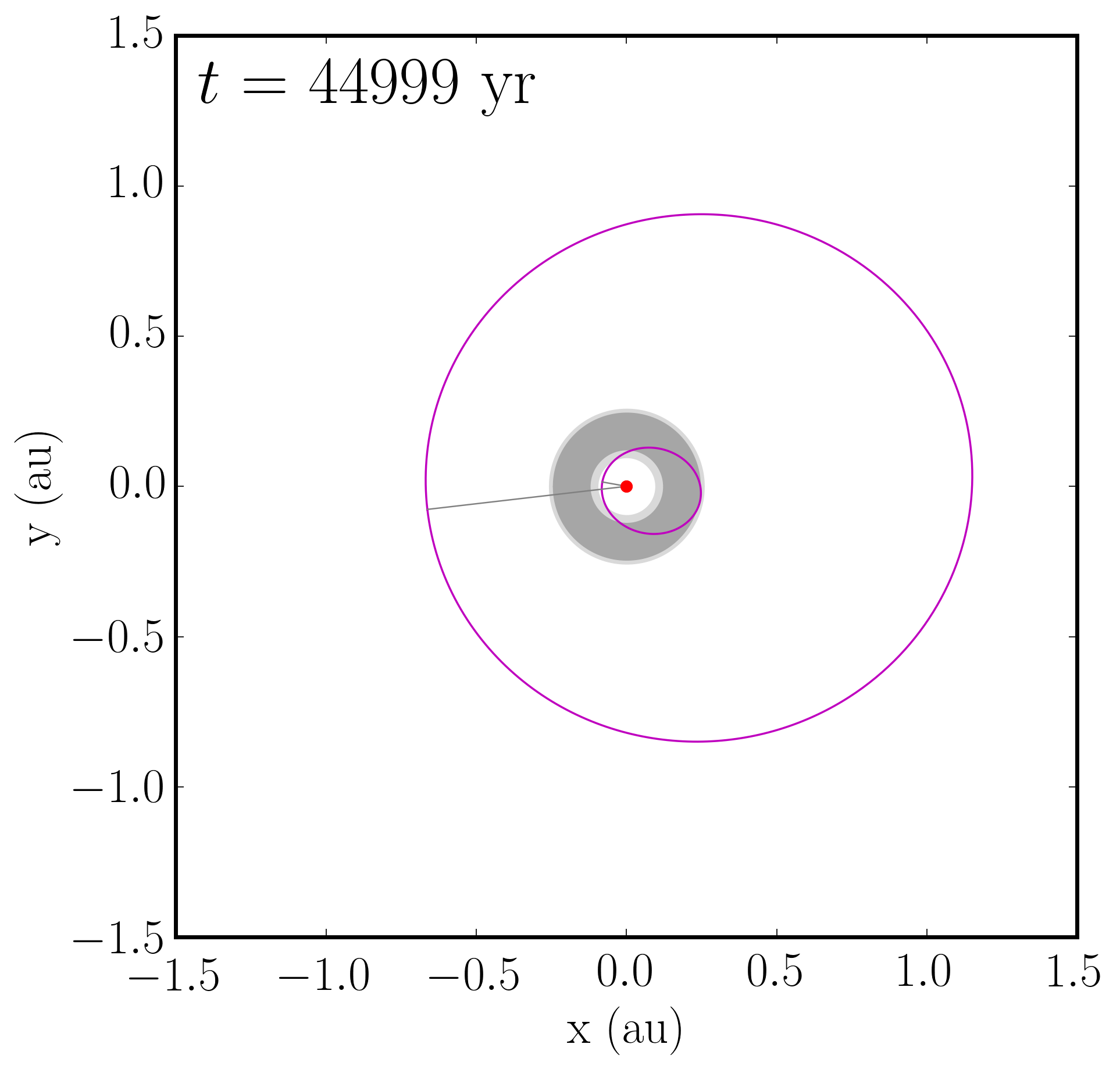} \\
    \includegraphics[width=5cm]{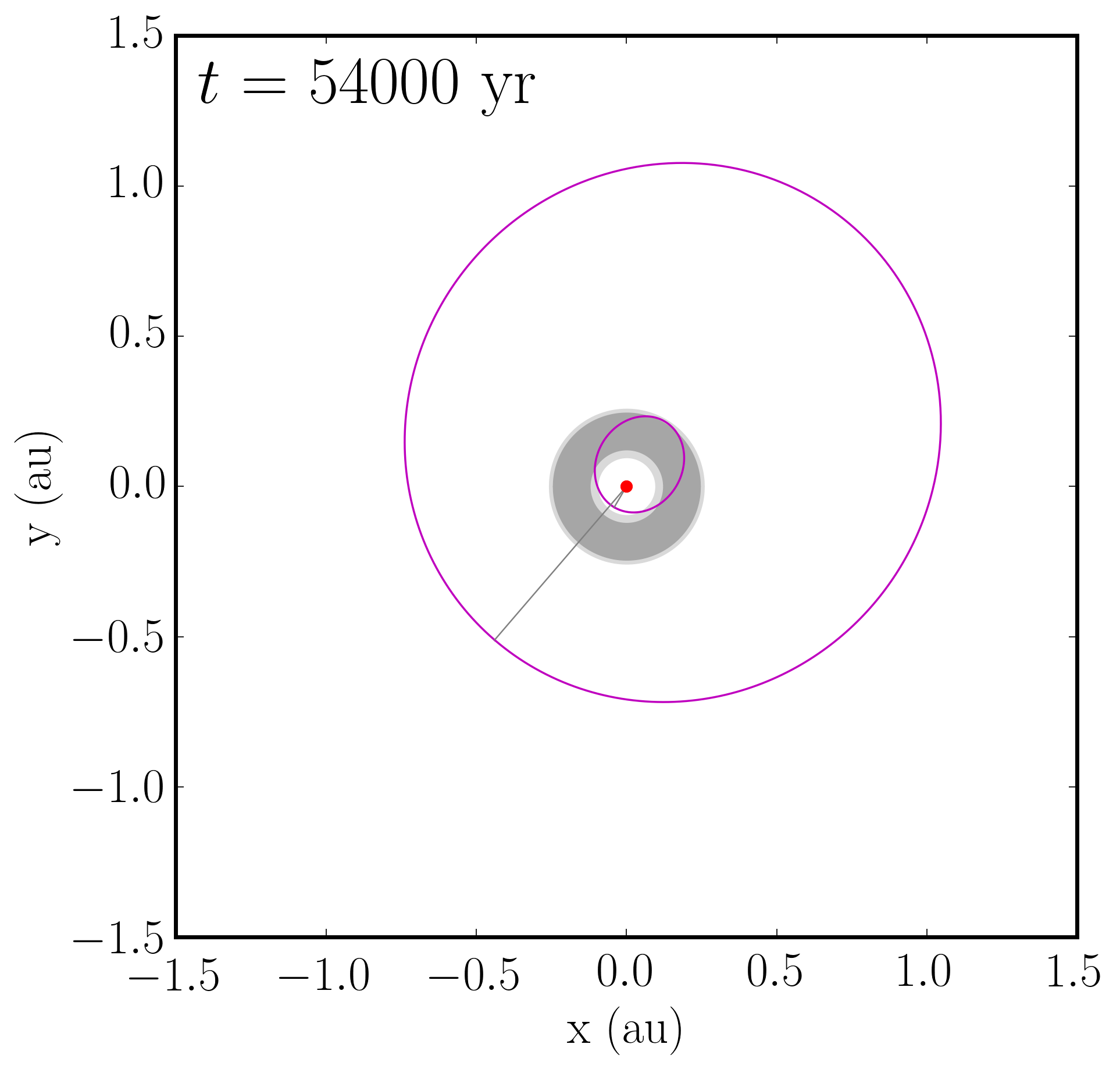} 
    \includegraphics[width=5cm]{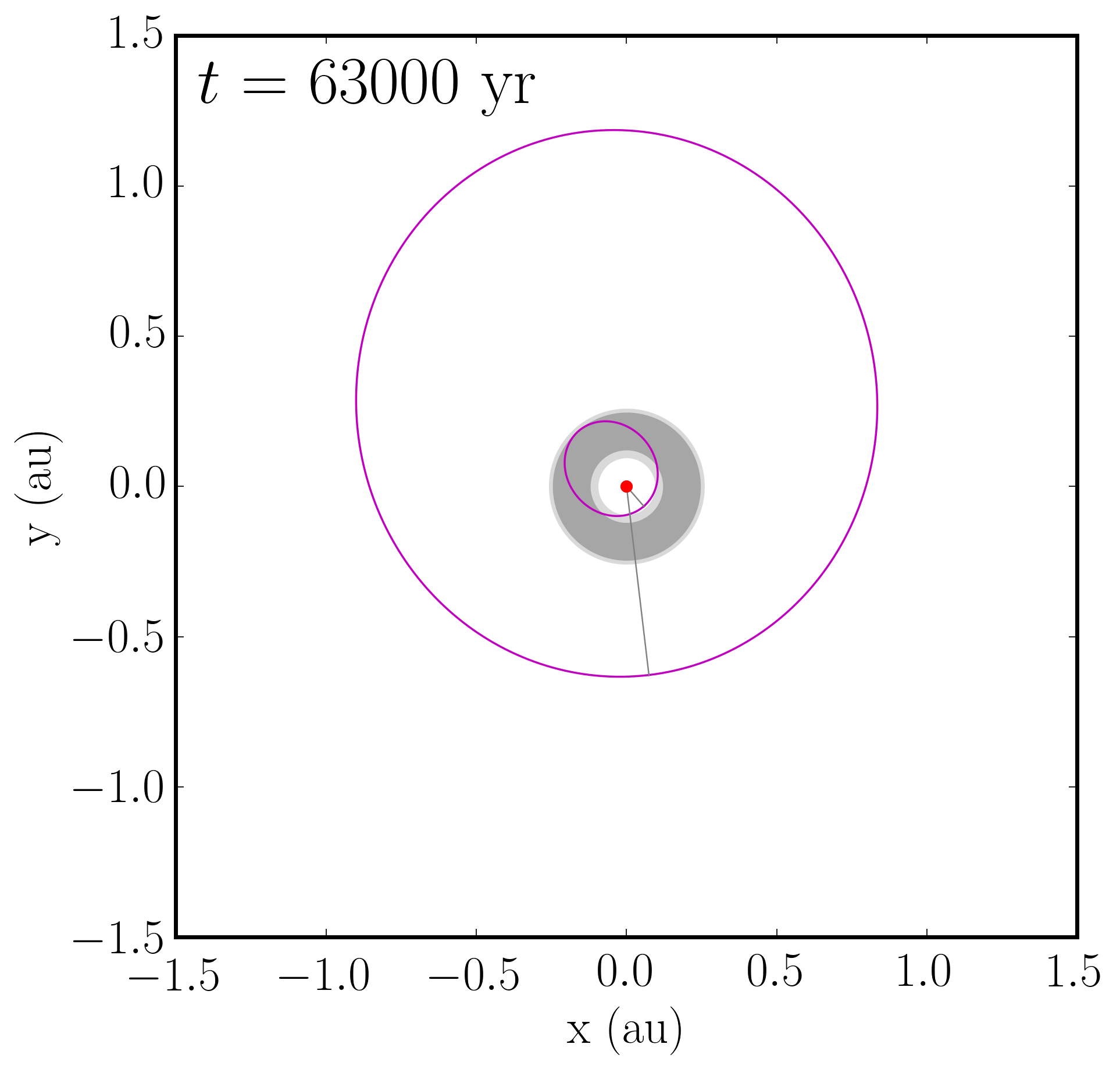} 
    \includegraphics[width=5cm]{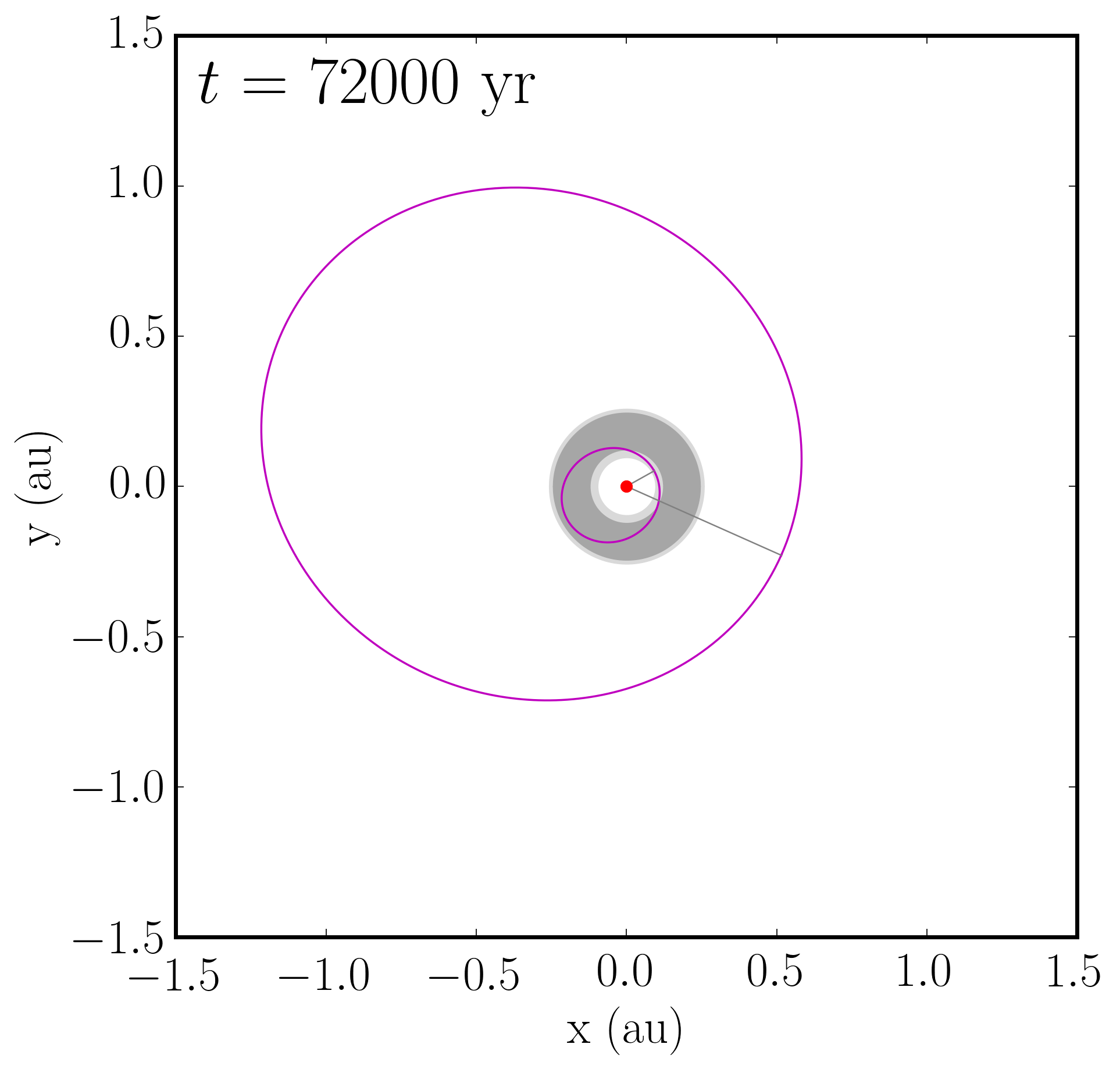} \\
\end{array} $

    \caption{Face-on representation of the orbital evolution
of the best fit of the GJ\,1148 system with a time step of 9000\,yr.
covering a full secular cycle of the planetary eccentricities, 
which has a period of $\approx$ 72500\,yr. 
Ellipses illustrate the planetary orbits with the line connecting 
the central star (red dot) and the planetary argument of
periastron $\varpi$. The timescale for orbital precession is $\approx$ 52000\,yr,
thus shorter than the secular period of the system.
The gray shaded disk represents the optimistic habitable zone around the  GJ\,1148 M dwarf.  Most of the time GJ\,1148 b orbits mostly within the HZ.
}
    \label{face_on} 
\end{figure*}

\subsection{Stability of the mutually inclined configurations}
\label{Sec4.2}

We now study the mutual inclination limits of the GJ\,1148 system by allowing the planetary
inclinations $i_{\rm b}$ and $i_{\rm c}$  and the difference of the line of nodes $\Delta\Omega$ = $\Omega_{\rm b}$ - $\Omega_{\rm c}$ to 
vary together with the remaining parameters in the dynamical modeling.
Initially we tried a simplex optimization with $i_{\rm b}$, $i_{\rm c}$, and $\Delta\Omega$ being free,
but this did not yield a significantly different best fit when compared to the co-planar edge-on model. 
The likely reason is that there is no sharp $-\ln\mathcal{L}$ minimum in the  $i_{\rm b}$, $i_{\rm c}$, and $\Delta\Omega$  space to 
which the simplex algorithm could to converge. 
This is expected since constraining 
the mutual inclinations from RV data requires strong interactions between the planets over the temporal baseline of the observations.
The best example is the other massive pair around an M dwarf; for the GJ\,876 system, \citet{Rivera2010}, \citet{Nelson2016},
and \citet{Millholland2018} 
successfully measured the orbital inclinations from RV data.
For GJ\,876, the timescales of the perturbations are shorter than the observational data time span, and thus it is possible to 
constrain the mutual inclination.
As we showed in \autoref{Sec5.1}, the timescales of the perturbations in the GJ\,1148 system are 
much longer, and thus it is hardly possible to constrain the inclinations by modeling the RV data.

\begin{table}
    \centering   
    \caption{Stable posterior $1\sigma$ confidence levels for the dynamical masses, semi-major axes, and  inclinations of the GJ\,1148 system derived from a mutually inclined dynamical modeling of the RV data via MCMC.
The posterior confidence levels for the other planetary parameters are consistent with those obtained 
for the coplanar edge-on case (see \autoref{table:2}), and are thus  not shown here.}   
    \label{table:mut_incl}      

    \begin{tabular}{@{}lcc@{}}
    \hline\hline  \noalign{\vskip 0.7mm}      
    Parameter &     GJ\,1148 b & GJ\,1148  c\\
    \hline \noalign{\vskip 0.7mm} 
    $m$ [$M_{\rm Jup}$]                          &    [0.300; 0.432]       &   [0.205; 0.313]     \\\noalign{\vskip 0.7mm} 
    $a$ [au]                          &    [0.165667; 0.165689] &   [0.908689; 0.911094] \\\noalign{\vskip 0.7mm} 
    $i$ [deg]                     &    [57.36; 122.97]      &   [58.18;119.64] \\\noalign{\vskip 0.7mm} 
    $\Omega$[deg]                     &    0 (fixed)            &   [0.0; 248.43]  \\\noalign{\vskip 0.7mm} 
    $\Delta i^b$ [deg]              &                         &   [0.0; 60.0 ]      \\\noalign{\vskip 0.7mm}

    \hline %\noalign{\vskip 0.7mm} 
    \end{tabular}  
    \tablefoot{
        %$a$ -- a prior applied, 
        $a$ -- small possibility for retrograde orbits with 130$^\circ < \Delta i < 180^\circ$, see \autoref{mcmc_mut_incl}
    }
\end{table}

Therefore, instead of optimizing the  $-\ln\mathcal{L}$ via the simplex algorithm, we simply ran an MCMC 
test starting from the best coplanar edge-on fit and we allowed $emcee$ \citep{emcee} to explore
the parameter space consistent with the combined HIRES and CARMENES RVs. 
For all parameters in this test we adopted flat priors with rather relaxed boundaries;
 the only exception was $\Omega_b$, which we fixed at 0$^\circ$ because the solutions are degenerate in this parameter. 
The mutual inclinations of the MCMC configurations are calculated as
\begin{equation}
    \begin{split}
    \Delta i = \arccos(\cos i_{\rm b}\cos i_{\rm c} + \sin i_{\rm b} \sin i_{\rm b} \cos\Delta\Omega)
    \end{split}
    \label{eq:deltai}
,\end{equation}
where $\Delta\Omega$ = $\Omega_b$ -- $\Omega_c$ is the difference between the 
orbital line of nodes. Since only $\Delta\Omega$ is important for deriving $\Delta i$, $\Omega_b$ can remain fixed at 0$^\circ$.

\Autoref{mcmc_mut_incl} shows the results from our dynamical MCMC allowing for mutual inclinations. 
The samples that survived 10 Myr without leading to a strong chaotic behavior
are found only in prograde configurations in the range   0$^\circ < \Delta i \lesssim 60^\circ$ 
or, less likely but still possible, on a retrograde  mutually inclined configuration in the range 130$^\circ \lesssim \Delta i < 180^\circ$.
With these results, we cannot firmly constrain the mutual inclinations between the planets,
but we can derive limits on their dynamical masses.
Most of the configurations in the MCMC posterior distribution 
can be found in the range between 57.36$^\circ <   i_b < 122.97^\circ$, and between 58.18$^\circ <   i_c < 119.64^\circ$.
The $\sin i$ factor for these inclinations is $\ga$ 0.85, which leads to maximum dynamical planetary masses of
$m_b$ = 0.432 $M_{\rm Jup}$ and  $m_c$ = 0.313 $M_{\rm Jup}$.
\autoref{table:mut_incl} summarizes the stable posterior $1\sigma$ confidence levels of the 
dynamical masses and semi-major axes constrained by the possible orbital inclinations of the GJ\,1148 system.

\begin{figure*}[tp]
    \centering
    \includegraphics[width=18cm]{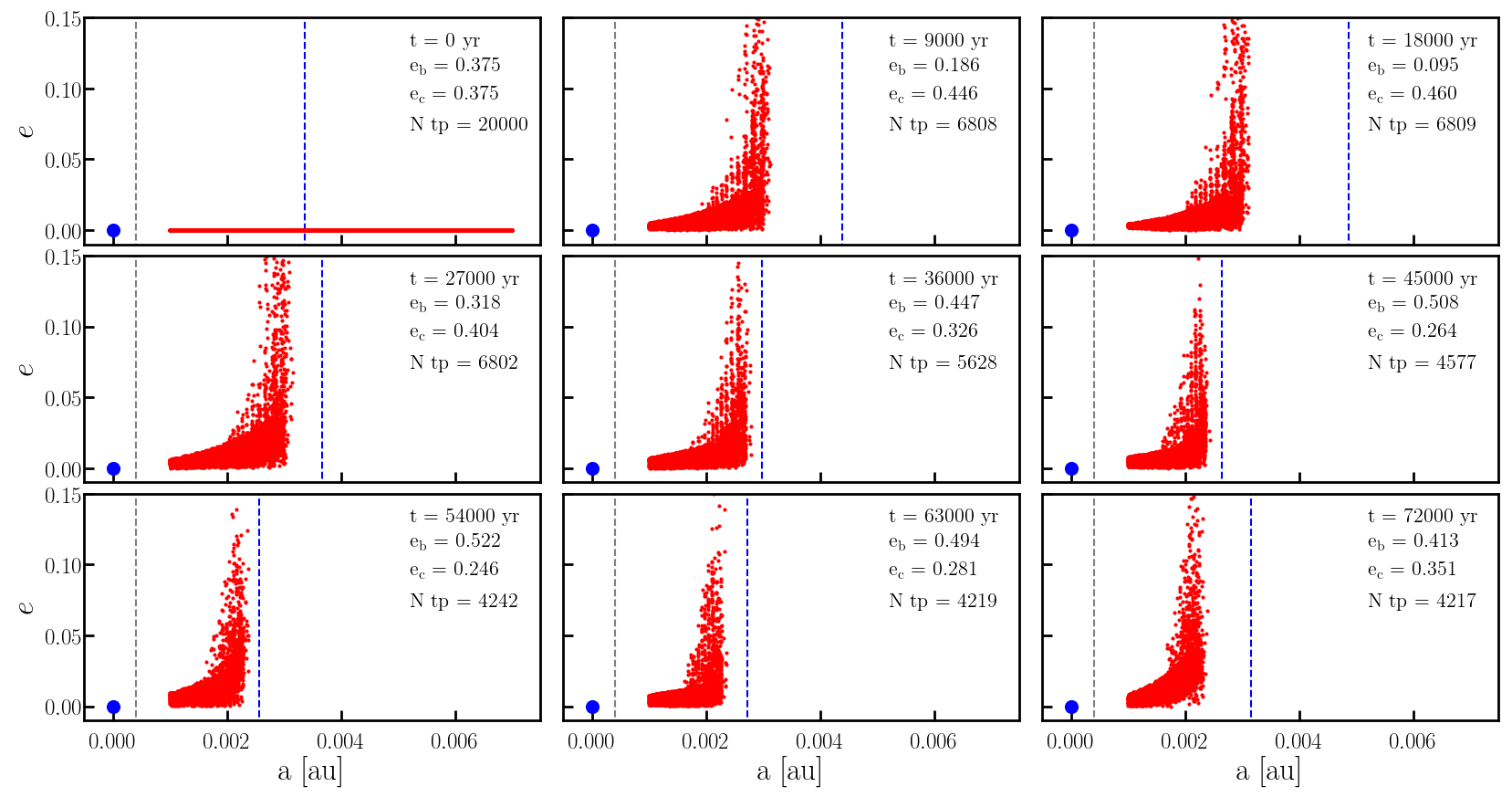} 

    \caption{
Evolution of mass-less test particles (i.e., exomoons) around GJ\,1148 b, under the gravitational perturbation of the outer planet.
Each panel tracks the evolution of the test particles with a step of 9\,000 yr to match the snap-shot evolution of the massive bodies in the GJ\,1148 system shown in \autoref{face_on}. Shown are the position of GJ\,1148 b (blue dot), the planetary Roche limit (gray dashed line), and the planetary $0.5R_{\rm Hill}$ (blue dashed line), which scales with (1-$e_b$) due to the dynamical perturbations of GJ\,1148 c. See text for details. 
    }
    \label{fit_tp} 
\end{figure*}

 \subsection{Prospects of habitable exomoons around GJ\,1148 b  }
\label{Sec4.5}

The GJ\,1148 system is very interesting, particularly because  the inner giant GJ\,1148 b 
resides inside the inner edge of the optimistic HZ.
\Autoref{face_on} shows a face-on representation of the secular cycle of the GJ\,1148 system.
Each panel of \autoref{face_on} shows a time step of 9\,000 yr of the evolution, covering a full 
cycle of the planetary eccentricities  which has a period of $\approx$ 72\,500\,yr and a full cycle of the 
orbital precession of the system, which has a period of $\approx$ 52\,000\,yr. 
The two gray overlapping annuli in the panels shows the optimistic and conservative HZ range around the GJ\,1148 M dwarf.
For the HZ estimate we made use of the {\em Habitable Zone Calculator}\footnote{\url{https://depts.washington.edu/naivpl/content/hz-calculator}},
which relies on the work by \citet{Kopparapu2013}, who provide HZ estimates 
around MS stars with effective temperatures in the range of 2600\,K $< T_{\rm eff} <$ 7200\,K.
From \autoref{face_on} it is evident that GJ\,1148 b is orbiting within the optimistic HZ, but it leaves the conservative HZ temporarily when its orbit is most eccentric.

While  GJ\,1148 b is unlikely to be hospitable to life as we know it, the question arises of whether  it could host habitable moons. 
We refer the reader to \citet{Zollinger2017} 
for more detailed description of exomoons around Saturn-mass planets orbiting M-dwarf stars and to
\citet{HellerBarnes2013}, \citet{Heller2014}, and \citet{HellerBArnes2015}
for further discussion on the habitability of exomoons. 
Here we focus in more detail on the GJ\,1148 b exomoon hypothesis.\looseness=-8

It is reasonable to assume that GJ\,1148 b was formed farther out, beyond the protoplanetary ice-line around the GJ\,1148 M dwarf 
where it can grow massive enough to become a Saturn-mass planet, and migrated inwards to more temperate orbits. 
Given our best local example, Saturn, it is also reasonable to assume that GJ\,1148 b has icy exomoons, or had them in the past.  
In the HZ, such exomoon bodies may not be icy anymore, but potentially habitable, ocean-like worlds.

The formation of moons around planets is similar to the formation of planets in protoplanetary 
disks. Due to type I migration torques, the regular moons, which are close to the 
planet, are thought to have formed in situ during the late stages in a starved circumplanetary 
disk before its dispersal \citep{canup2002}. Conversely, the irregular satellites, which usually orbit 
farther out, are smaller in size, and have eccentric and inclined orbits, are 
thought to have been dynamically captured (see \citealp{nader2007} for a review).

Additionally, exomoons will have some benefits with respect to planetary bodies orbiting around an M dwarf inside the HZ.
For example, an M-dwarf HZ planet will be tidally locked to the star, with only one side facing the star. It is still debatable whether such tidally locked planets could host life 
due to the large temperature contrast between the day and night sides of the planet, but most recent research suggests tidally locked planets can maintain surface liquid water \citep{Joshi1997,Pierrehumbert2011,Yang2013,DelGenio2019}. On the other hand, exomoons will be tidally locked to the giant planet, not to the star. This may allow more 
uniform irradiation of the exomoon while it orbits around its planetary host,  thus overcoming the problems the HZ planets may have.
In addition, HZ planets around M dwarfs may experience intensive stellar flares typical for M-dwarf stars. 
The massive  Saturn-mass planet may, however, have a strong magnetic field, which could act as an effective shield 
protecting the habitable moon from stellar activity \citep{HellerZuluaga2013}.
Of course, these assumptions are highly speculative, while many physical 
considerations must be taken into account to test how feasible these scenarios are. 
From the dynamical perspective, long-term stability and tidally induced satellite orbit decay can limit the 
existence of habitable exomoons around GJ\,1148 b \citep{BarnesOBrien2002,SasakiBarnes2014}, and we  look further into these possible limitations.

\begin{figure*}[tp]
    \begin{center}$
        \begin{array}{cc} 
            \includegraphics[width=18cm]{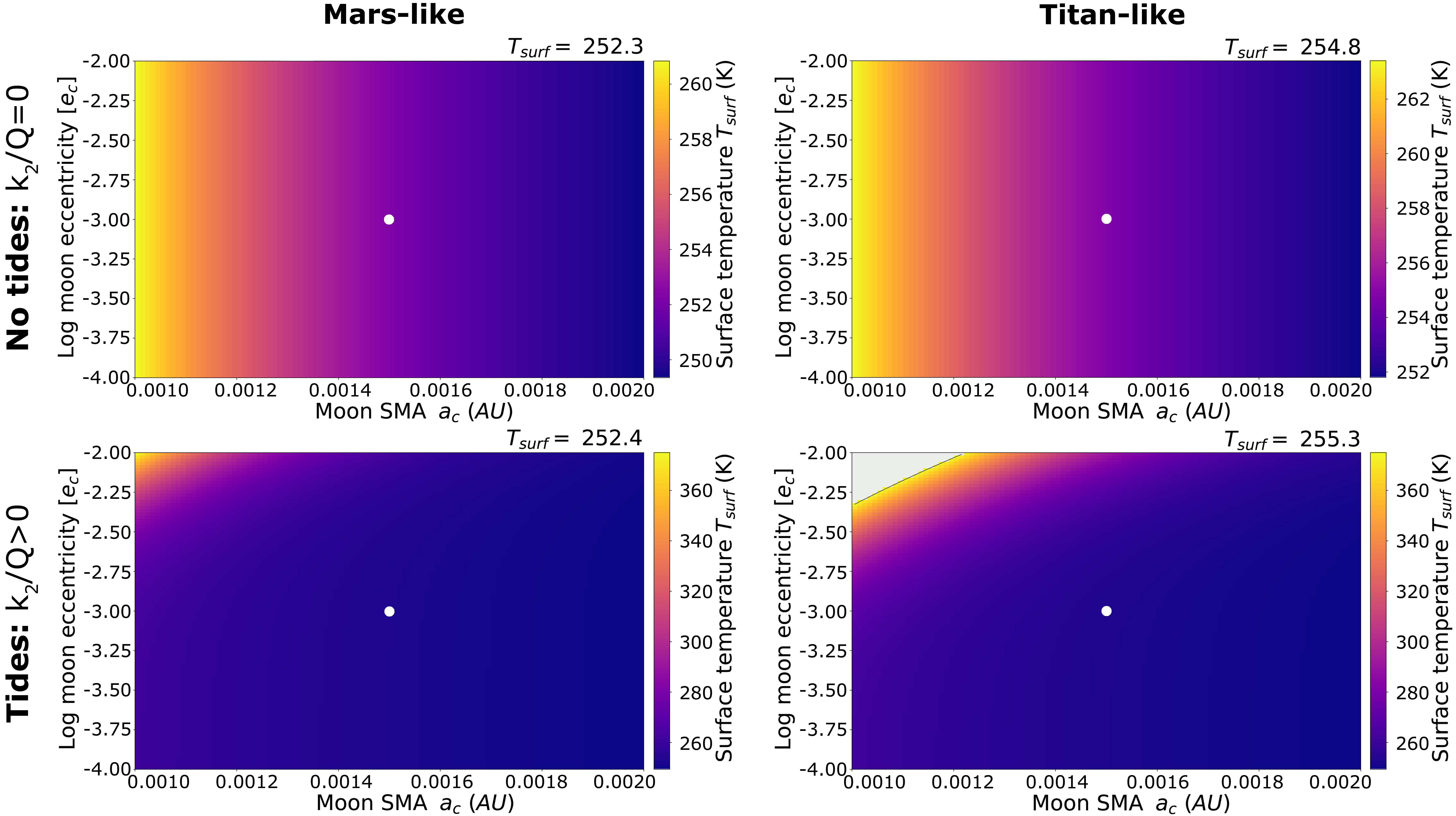} 
        \end{array} $
    \end{center}
    \caption{Surface temperatures given varying eccentricities and semi-major axes for both a Mars-like (left) and Titan-like (right) moon around GJ\,1148 b. The color bars are not identical for each plot. White dots denote the fiducial values; light gray regions indicate surface temperatures in excess of the boiling point of water at 1 atmospheric pressure.}
    \label{surftemps} 
\end{figure*}

\subsubsection{Stability border}
\label{sect451}

The natural scale for the Hill stability of an exomoon is the Hill radius, which for GJ\,1148 b is

\begin{equation}
    R_{\rm Hill} = q_b\left(\frac{m_b}{3M_{\star}}\right)^{1/3} \approx 0.006\,{\rm au},
    \label{eq:hill}
\end{equation}

\noindent
where $q_b$ = $a_b(1-e_b)$ is the periastron distance,
$m_b$ is the planetary mass of GJ\,1148 b,  and $M_{\star}$ is the stellar mass of GJ\,1148.
 \cite{Grishin2017}  have found an analytic fit for the stability border as a function of arbitrary mutual inclination. The stability border is around $0.5 R_{\rm Hill}$ for prograde coplanar configurations; decreases to $0.4 R_{\rm Hill}$ for highly inclined configurations with $i \sim 90^{\circ}$; and increases to $0.7$--$1 R_{\rm Hill}$ for retrograde coplanar configurations, whose  orbits at $\ga 0.6 R_{\rm Hill}$ are chaotic and have high eccentricities. 
Retrograde configurations have long been known to be more stable than prograde ones \citep{henon1970, hb91},
but are also less likely to occur; thus, we do not study retrograde exomoon orbits and focus only on the co-planar prograde case.

For the GJ\,1148 system, the outer planet GJ\,1148 c could further reduce the stability border, because GJ\,1148 c secularly drives the eccentricity and changes the Hill radius of GJ\,1148 b. The Hill stability of exomoons was studied in a different context by \citet{Hamers2018} around circumbinary planets, which also further reduced the fraction of stable orbits, but for different reasons involving resonances between the inner binary star and the orbit of the exomoon.

In order to explore the stability of exomoons around GJ\,1148 b, we performed N-body integrations of test particles around the temperate planet GJ\,1148~b.
In this test, we randomly placed 20\,000 test particles with semi-major axes in the range 0.001--0.007 au on initially circular orbits.
The semi-major axis range was chosen carefully to avoid the planetary Roche limit, inside of which the exomoon cannot exist, 
and to be inside the planetary Hill radius.
The integrations were done using the same Wisdom-Holman algorithm \citep[][]{Wisdom1991}
that we used in our long-term stability tests based on the MCMC analysis, but modified to handle the evolution of additional 
test particles in the Jacobi coordinate system \citep{Lee2003}. 
We inverted the hierarchy of the system, making the GJ\,1148 b planet the central body, orbited by the non-mutually-interacting massless test particles, the 
GJ\,1148 stellar body, and the  outer planetary body GJ\,1148 c, in that order. 
The initial parameters of the massive bodies were adopted from the best coplanar edge-on dynamical model.
Since the integrations were done in Jacobi frame, the evolution of the three massive bodies will be  
the same as shown in Figures~\ref{orb_evol},~\ref{oct_evol}, and \ref{face_on} for the best fit, where the central body is the star. 
The time step chosen for this test is very small, only 0.01\,d, which is needed for the accurate N-body
integration of the closest test particles, which have periods of only $\sim$ 0.8\,d.

\Autoref{fit_tp} shows the results from the test particle simulations.
Each panel of \autoref{fit_tp} tracks the evolution of the test 
particles with a step of 8600 yr to match the snap-shot evolution of 
the massive bodies in the GJ\,1148 system shown in \autoref{face_on}. 
The blue dot indicates the position of GJ\,1148 b assuming that it is the central body, the gray dashed line marks the planetary Roche limit, 
while the blue dashed line marks the planetary 0.5R$_{\rm Hill}$.
The $0.5R_{\rm Hill}$ limit scales with (1-$e_b$) due to the dynamical perturbations of GJ\,1148 c, i.e., it is farther out when $e_b$ $\approx$ 0, and closer in when $e_b$ $>$ 0.
Test particles which fall outside the $0.5R_{\rm Hill}$ limit are quickly ejected, while those just inside are 
excited to large eccentricities, and also become unstable at some point during the integration.
To summarize, the stability border of the coplanar prograde 
case is around $\sim 0.0023$ au which is around $0.5{\rm min} \left\{R_{\rm Hill} \right\} $, where the minimum is obtained for the maximum value of $e_b$ during the secular period.
 
\begin{figure*}[tp]
    \centering 
    \includegraphics[width=.5\linewidth]{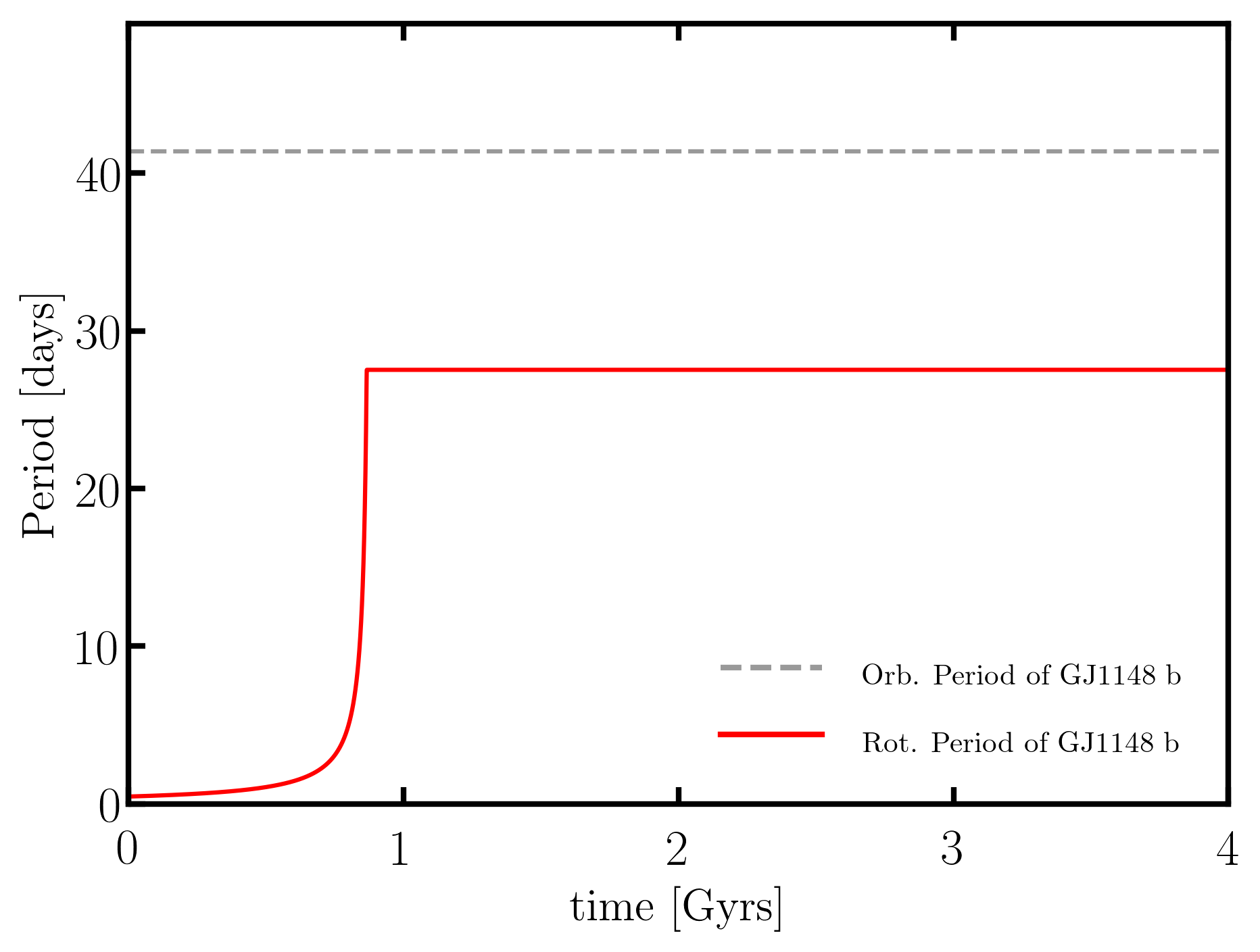}\includegraphics[width=.5\linewidth,height=7.0cm]{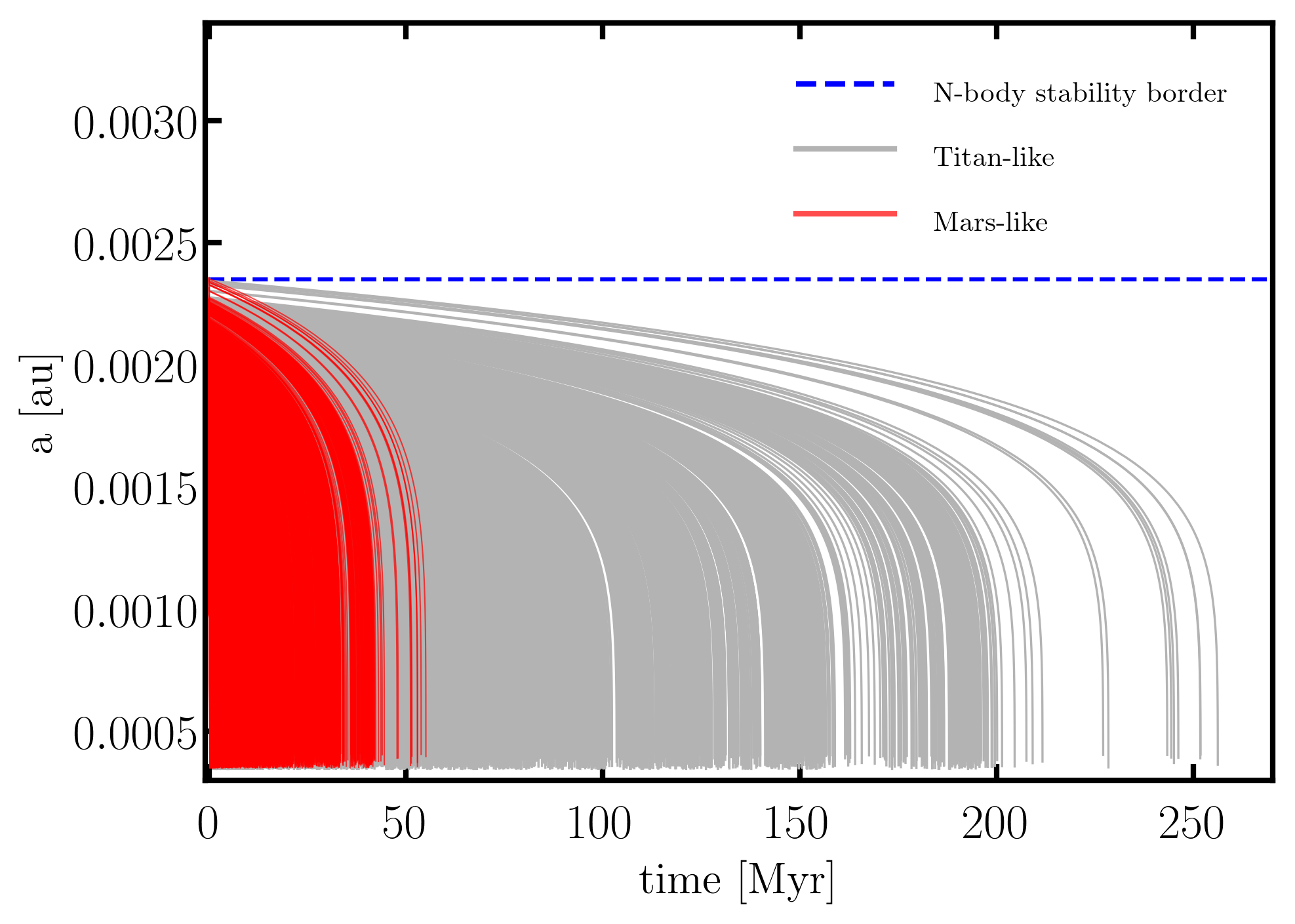} \\

    \caption{Spin evolution of GJ\,1148 b with adopted R = 9.44 $R_\oplus$, $k_{\rm 2}$ = 0.341, and Q = 18 000 due to star--planet tides (left) and tidal evolution of Mars-like 
    ($m$ = 0.107 $M_\oplus$, $R$ = 0.53 $R_\oplus$, $k_{\rm 2}$ = 0.14, $Q$ = 140) and Titan-like ($m$ = 0.0225 $M_\oplus$, $R$ = 0.404 $R_\oplus$, $k_{\rm 2}$ = 0.589, Q = 58.9) exomoons due to exomoon--planet tides (right).
    The initial rotational period of GJ\,1148 is P$_{\rm rot}$ = 11\,h (consistent with Saturn), 
    while of exomoons it is P$_{\rm rot}$ = 24\,h (consistent with Mars).
    Asymptotic rotation at $2/3$ of the orbital period is reached for GJ\,1148 b after $\sim$ 850\,Myr.
The Saturn-like planet therefore reaches P$_{\rm rot}$ = 27.5\,d,
which leads to strong orbital decay of the N-body stable exomoon orbits due to tides with the planet. The initial exomoon orbits are adopted from those that survived the N-body stability test in \autoref{sect451} (see also \autoref{fit_tp}).
}
    \label{tides} 
\end{figure*}

Most likely the larger exomoons will be well within the Hill sphere. Highly inclined exomoons may still be stable if they are within the Laplace radius \citep{tremaine09}

\begin{equation}
\begin{split}
r_L = \left(\frac{J_2 R_b^2 a_b^3 (1-e_b^2)^{3/2}m_b} {M_\star}\right)^{1/5} \end{split}
\label{eq:lap_rad}
,\end{equation}
where the inner quadrupole of the planet induced by its oblateness overtakes the outer quadrupole of the star. Here $J_2$ is a dimensionless coefficient that is related to the planet's oblateness, and $R_b$ is the radius of the planet. For Jupiter-mass planets,
the Laplace radius is a few dozen times  the planet's radius, $r_L \sim 30 R_b$. On the other hand, even moderate inclinations slightly outside the Laplace radius can result in unstable orbits either by the inner quadrupole \citep{tremaine09} or by chaotic evolution with an additional fourth body \citep{grish_lai18}. In either case the result can be the ultimate ejection or collision of the exomoon.

 \subsubsection{Tidal heating}
\label{452}

Tidal torques exerted on an exomoon by both the host planet and the host star
deform the exomoon. This dissipates heat in the  interior of the exomoon, 
which increases its surface temperature above the value resulting from stellar illumination alone. The amount of heat dissipated 
depends on the tidal efficiency factor $k_2/Q$: this value was determined to be $\approx 0.015$ for Io, and between $0.0026$ and $0.0127$ for Enceladus \citep{Nimmo2018}. 

We wrote an algorithm that computes the surface temperature of an exomoon of given physical and orbital 
characteristics around any particular exoplanet, then investigated how the surface temperature changes with 
semi-major axis, eccentricity, and tidal efficiency factor of the moon. 
We adopted two fiducial moons: one with the physical 
characteristics of Titan (Titan-like) and one with the physical characteristics of Mars (Mars-like). 
We varied the semi-major axis within the stability region (i.e., 0.001 AU $\leq a_m \leq$ 0.002 AU) with the 
base value at 0.0015 AU; we varied the eccentricity in the range 0.0001 $\leq e_m \leq$ 0.01, with the base value 
at 0.001 (one-fifth of the Enceladus eccentricity). 
We computed surface temperature with and without tidal heating; in the former case $k_2/Q = 0.01$ for the Titan-like 
moon and $k_2/Q = 0.001$ for the Mars-like moon \citep[literature values for Mars and other solar system bodies are listed in][]{Lainey2016}, 
while in the latter case, $k_2/Q$ was set to zero for both fiducial moons. 
The results are displayed in \autoref{surftemps}. 
White dots denote the fiducial values; we note that the color bars are {not} identical for all 
figures. Light gray regions indicate surface temperatures in excess of the boiling point of water at 1 atmospheric pressure.

In all cases, regardless of tidal heating, a moon at 0.0015 AU with eccentricity around
0.001 has a surface equilibrium temperature of around 255 K. Since this is above the snow line 
temperature, surface ice may evaporate to form an atmosphere. Hence, in terms of ambient temperature habitable exomoons around GJ\,1148 b are possible: if a sufficiently strong greenhouse effect takes place, surface conditions on a hypothetical exomoon GJ\,1148 b-I may allow  liquid water.

\subsubsection{Tidal torque drift due to planet--moon tides}

It is important to note that in our stability test, we neglected tidal interactions between the bodies; however, they are important in such a compact system and pose further obstacles for the existence of potentially habitable moons. % around GJ\,1148 b. 

There are two critical tidal timescales, which are relevant for exomoons around GJ\,1148 b: 
(i)~the spin-orbit synchronization timescale of GJ\,1148 b and 
(ii) the tidal torque drift timescale for the ``stable'' exomoon due to planet--moon tides.
To derive these timescales we use the \textsc{EqTide}\footnote{\url{https://github.com/RoryBarnes/EqTide}} code
\citep{Barnes2017}, which calculates the tidal evolution of two bodies based on models by \citet{Ferraz-Mello2008}.
To calculate (i), for the star--planet pair we adopted 
the mean osculating semi-major axis $a_{\rm b}$ = 0.166 au and mean eccentricity $e_{\rm b}$ = 0.361 for GJ 1148 b. 
Stellar physical parameters were adopted from \autoref{table.stars}, 
while for the planet the initial spin of $P_{\rm rot}$ = 11h was chosen, and the planetary radius set to 
$R_{\rm b}$ = 58\,230 km (consistent with Saturn). From \citet{Zollinger2017} we adopted Saturn-like physical parameters, 
such as love number $k_{\rm 2}$ = 0.341, and dissipation factor $Q$ = 18\,000 (see their Table 2). 
To calculate the tidal evolution in (ii), we adopted the stable test particles (i.e., exomoons) from \autoref{sect451},
and we adopted  Mars-like and Titan-like physical parameters 
(i.e., for Mars $m$ = 0.107 $M_\oplus$, $R$ = 0.53 $R_\oplus$, and for Titan $m$ = 0.0225 $M_\oplus$, $R$ = 0.404 $R_\oplus$), 
and for both an initial rotation period of 1\,d, which is consistent with the rotational period of Mars.  
For Mars-like moons we adopt $k_2$ = 0.14 from \citet{md99}, while
for Titan-like moons we adopt $k_2 = 0.589$ from \citet{Iess2012}. 
For consistency with the adopted $k_2$/Q values in \autoref{452},
we adopt $Q = 58.9$ for Titan, and $Q = 140$ for Mars.

\Autoref{tides} shows the results of our tidal evolution calculations. 
The left panel of \autoref{tides} shows the planetary rotational evolution of GJ\,1148 b due to star--planet tides.  
After $\sim$850\,Myr, GJ\,1148~b reaches a rotation period that is $2/3$ of 
the orbital period, and remains there with P$_{\rm rot}$ = 27.5\,d. During the integration 
the planetary semi-major axis and eccentricity are mostly unaffected.
An asymptotic rotation period that is shorter than synchronous and 2/3 of 
the orbital period is expected for $e_b \ga 0.24$ in the constant $Q$ tidal model \citep{Goldreich1966,Cheng2014}. 
The time for GJ\,1148~b to reach asymptotic rotation is inversely proportional to 
the initial $P_{\rm rot}$, as long as the initial $P_{\rm rot}$ is much less than $27.5\,$d, and it depends on the other parameters of GJ\,1148~b according to Equation (3) of \cite{Barnes2017} and Equation (15) of \cite{Cheng2014}.
The rotational period of GJ\,1148 b is thus very likely much longer than 
the orbital periods of the hypothetical exomoons, which could be 
dynamically stable only with orbital periods between 0.7 to 2\,d. 
The right panel of \autoref{tides} shows that the longer rotational period of GJ\,1148 b 
(P$_{\rm rot}$ = 27.5\,d) leads to strong orbital decay of the stable exomoon orbits due 
to tidal interactions with the planet. An exomoon eventually 
reaches the Roche limit where it is tidally disrupted by the gas giant.  
Not even one hypothetical ``stable'' exomoon  in the context of \autoref{sect451} had survived this test.
The maximum time a Mars-like exomoon could survive is $\sim$ 55 M\,yr,
while for Titan-like moons the maximum survival time is longer, $\sim$ 255 M\,yr. 
The latter is longer by roughly the mass ratio of Mars to Titan, which can 
be understood from Equation (2) of \cite{Barnes2017} and Equation (16) of \cite{Cheng2014}. 
These timescales are  optimistic since the orbital decay would start before the planet reaches the asymptotic spin state.
In both cases the survival times are much shorter than the age of the system. 
Therefore, given the relatively fast orbital decay in the small stable region 
around the planet, we conclude that exomoons around GJ\,1148 b are unlikely to exist.

It should be noted that \cite{MartinezRodriguez2019} also considered the tidal evolution of a
hypothetical exomoon of GJ 1148 b and reported a timescale for orbital
migration larger than 14\,Gyr, a dramatically different result. The
source of the discrepancy is that they assumed a much longer orbital
period for the moon than we do (8 days versus $< 2$ days), which changes the
tidal evolution by orders of magnitude. 
In addition, they did not calculate
the stability limit for the moon's orbit.

\section{Conclusions and discussion}
\label{Sec6}

We present an updated orbital solution and dynamical analysis of the GJ\,1148 M-dwarf 
planetary system based on new CARMENES optical RV data. 
We also present CARMENES near-IR RVs for GJ\,1148, which are in excellent agreement with 
the available optical data from HIRES and CARMENES.
The GJ\,1148 system is in a peculiar configuration consistent with a pair of Saturn-mass gas giants, which 
exhibit large eccentricity oscillations over a secular timescale of $\sim$ 67\,000 years. 
Even so, the system is well separated and remains stable for a large set of coplanar and mutually inclined configurations.  
Characteristic for the GJ\,1148 system is the evident apsidal alignment 
due to the large eccentricities, but overall we did not find any
high-order mean-motion resonant commensurabilities in which the system could be trapped.
It is possible that the observed high eccentricities are the result of planet-planet 
scattering events, which could possibly have occurred in the early phases of the history of the system.

We find that the existence of potentially habitable exomoons around GJ\,1148 b 
is unlikely due to the narrow stability region around the close and eccentric planet.
Additionally, the tidally driven orbit decay timescales are much shorter than the age of the GJ\,1148 system. 
When considering both disturbances by an outer planet in a system with 
significant eccentricities  and tidal effects, there may be no space left for moons.
Even if exomoons had resided  in the stable region around the inner planet in the early phase 
of the history of the system, they would eventually  spiral towards the planet.

{\em TESS} will observe GJ\,1148 in Sector 22 from 18 February 2020 to 18 March 2020. 
Given the stellar radius estimated by \citet{Schweitzer2019}
and the planetary orbital eccentricities and semi-major axes we obtain in this work, we find a geometric
transit probability of $\sim$ 1\% for GJ\,1148 b and $\sim$ 0.1\% for GJ\,1148 c \citep[see, e.g.,][]{Barnes2007}. 
Given the short duration ($\sim$27 days) of the TESS photometric observations, 
we can expect to detect at most a single transit event for each planet.
Based on the posterior distribution of the orbital parameters from this work, we predict that 
a potential transit of GJ\,1148 b will occur on 13 February 2020 
around 11:30:00.0 UT $\pm$ 4h, and the next one will be on 25 March 2020
$\sim$ 20:40:00.0 UT $\pm$ 4h. So both are unfortunately outside the expected {\em TESS} window.
For GJ\,1148 c, the situation is even worse with the closest potential transit 
calculated to occur 84 $\pm$ 12\,d before {\em TESS}  starts the observations of Sector 22.
Therefore, there is no chance to observe potential transits of GJ\,1148 b \& c with {\em TESS}.
Nevertheless, {\em TESS} could reveal shorter period, low-mass transiting planets whose RV signature 
could be below the RV jitter we record with HIRES and CARMENES, and thus remain undetected 
by the Doppler method.

\begin{acknowledgements}
  CARMENES is an instrument for the Centro Astron\'omico Hispano-Alem\'an de
  Calar Alto (CAHA, Almer\'{\i}a, Spain). 
  CARMENES is funded by the German Max-Planck-Gesellschaft (MPG), 
  the Spanish Consejo Superior de Investigaciones Cient\'{\i}ficas (CSIC),
  the European Union through FEDER/ERF FICTS-2011-02 funds, 
  and the members of the CARMENES Consortium 
  (Max-Planck-Institut f\"ur Astronomie,
  Instituto de Astrof\'{\i}sica de Andaluc\'{\i}a,
  Landessternwarte K\"onigstuhl,
  Institut de Ci\`encies de l'Espai,
  Insitut f\"ur Astrophysik G\"ottingen,
  Universidad Complutense de Madrid,
  Th\"uringer Landessternwarte Tautenburg,
  Instituto de Astrof\'{\i}sica de Canarias,
  Hamburger Sternwarte,
  Centro de Astrobiolog\'{\i}a and
  Centro Astron\'omico Hispano-Alem\'an), 
  with additional contributions by the Spanish Ministry of Economy,
  the German Science Foundation (DFG),
  the Klaus Tschira Stiftung,
  the states of Baden-W\"urttemberg and Niedersachsen,
  the DFG Research Unit FOR2544 "Blue Planets around Red Stars, project RE 2694/4-1,
  and by the Junta de Andaluc\'{\i}a. 
We acknowledge financial support from the Agencia Estatal de Investigaci\'on of the Ministerio de Ciencia, Innovaci\'on y Universidades and the European FEDER/ERF funds through projects AYA2016-79425-C3-1/2/3-P, AYA2015-69350-C3-2-P, ESP2017-87676-C5-2-R, ESP2017-87143-R, and the Centre of Excellence ''Severo Ochoa'' and ''María de Maeztu'' awards to the Instituto de Astrof\'{\i}sica de Canarias (SEV-2015-0548), Instituto de Astrof\'{\i}sica de Andalucía (SEV-2017-0709), and Centro de Astrobiolog\'{\i}a (MDM-2017-0737)
This work has made use of data from the European Space Agency (ESA)
mission {\it Gaia} (\url{https://www.cosmos.esa.int/gaia}), processed by
the {\it Gaia} Data Processing and Analysis Consortium (DPAC,
\url{https://www.cosmos.esa.int/web/gaia/dpac/consortium}). Funding
for the DPAC has been provided by national institutions, in particular
the institutions participating in the {\it Gaia} Multilateral Agreement. M.H.L. and K.H.W. were supported in part by Hong Kong RGC grant HKU 17305618. R.B. was
supported by the NASA Virtual Planetary Laboratory Team through Grant
Number 80NSSC18K0829.
T.H. acknowledges support from the European Research Council under the
Horizon 2020 Framework Program via the ERC Advanced Grant Origins 83 24 28.
T.T.\ thanks Alexandre Correia and Laetitia Rodet for helpful discussions.
We thank the anonymous referee for the excellent comments that helped to improve the quality of this paper.

\end{acknowledgements}

\bibliographystyle{aa}

\bibliography{carm_bib}

\clearpage

\begin{appendix} %First online appendix

\label{appendix}

 \setcounter{table}{0}
\renewcommand{\thetable}{A\arabic{table}}

\setcounter{figure}{0}
\renewcommand{\thefigure}{A\arabic{figure}}

\begin{figure*}[ht]
    \includegraphics[width=18cm]{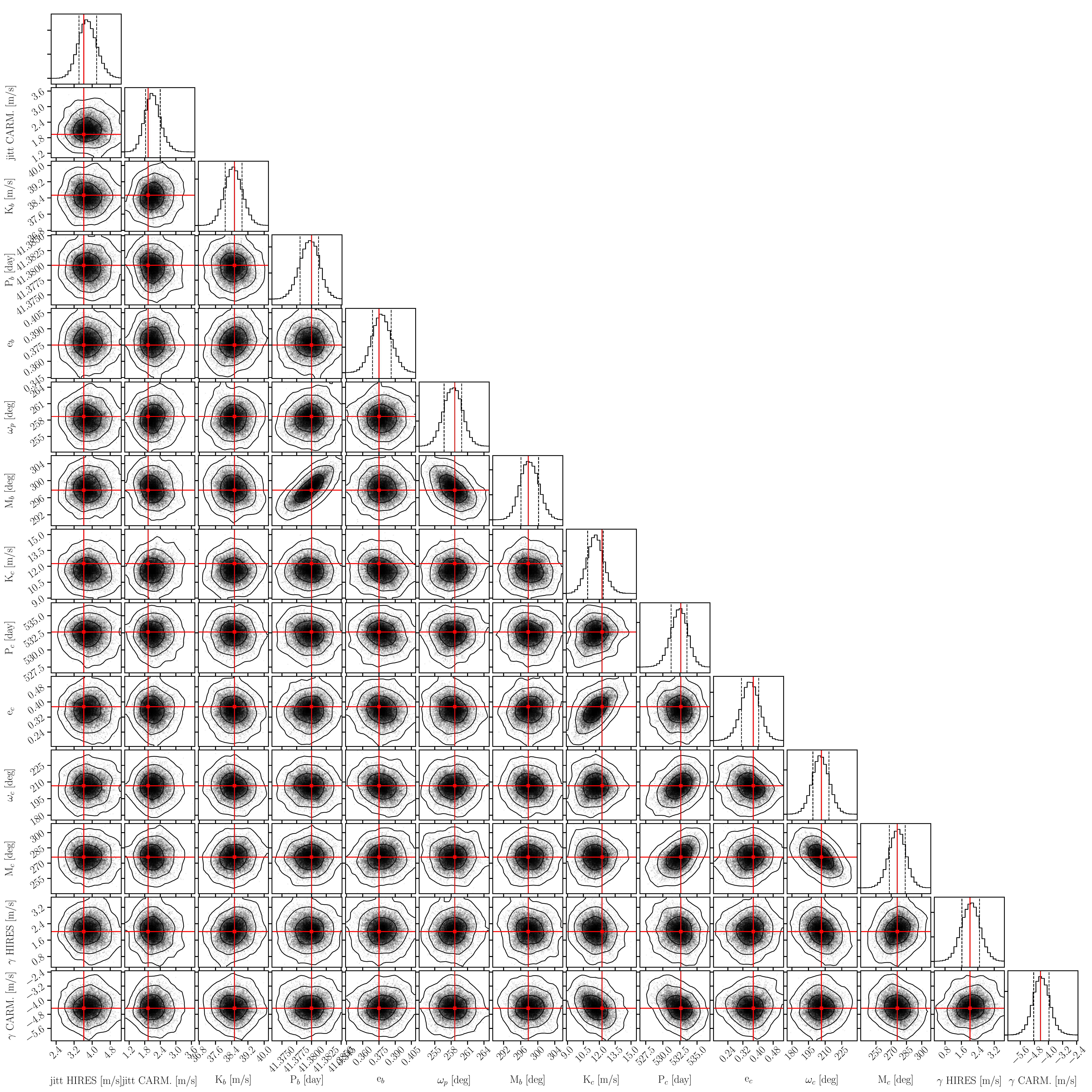}  
    \caption{MCMC distribution of orbital parameters 
consistent with the HIRES and CARMENES RV data of GJ\,1148 
assuming a coplanar, edge-on, and prograde 
two-planet system fitted with a self-consistent dynamical model.
The position of the best dynamical fit is marked with red grid lines.  
The black contours on the 2D panels represent the 1, 2, and 3$\sigma$ confidence level of the overall MCMC samples.
The top panels of every column represent the 1D histogram distribution of each parameter. All samples are stable for 10 Myr. 
    }
    \label{FigApp_mcmc}
\end{figure*}

\begin{table*}
\caption{CARMENES optical Doppler measurements and activity index measurements of GJ\,1148  } 
\label{tab:CARM_VIS} 

\centering  

\begin{tabular}{c c c c c c c c c c c c c c } 

\hline\hline    
\noalign{\vskip 0.5mm}

Epoch [JD] & RV [m\,s$^{-1}$] & $\sigma_{RV}$ [m\,s$^{-1}$]  &  H$\alpha$ & $\sigma_{H\alpha}$  &  dLW & $\sigma_{\rm dLW}$  &  CRX & $\sigma_{\rm CRX}$  & cairt & $\sigma_{\rm cairt}$  & \\  

\hline     
\noalign{\vskip 0.5mm}    

2457414.660   &   -37.064   &    1.325 &   0.910   &    0.003 &   -7.840   &    2.005 &   3.246   &    11.773 &   0.679   &    0.002 &      \\ 
2457419.695   &   -41.821   &    1.177 &   0.905   &    0.002 &   -11.100   &    1.375 &   2.312   &    9.812 &   0.679   &    0.002 &      \\ 
2457476.534   &   13.698   &    1.359 &   0.924   &    0.002 &   -6.067   &    0.936 &   18.732   &    9.137 &   0.679   &    0.002 &      \\ 
2457497.406   &   -46.393   &    2.374 &   0.905   &    0.002 &   -4.342   &    1.163 &   14.770   &    8.641 &   0.676   &    0.002 &      \\ 
2457510.440   &   22.313   &    1.466 &   0.909   &    0.003 &   -11.312   &    1.284 &   4.953   &    12.231 &   0.672   &    0.002 &      \\ 
2457529.389   &   -21.802   &    1.534 &   0.921   &    0.002 &   -7.669   &    1.747 &   1.987   &    10.089 &   0.671   &    0.002 &      \\ 
2457754.710   &   11.900   &    1.425 &   0.889   &    0.005 &   22.055   &    3.561 &   -69.990   &    19.556 &   0.668   &    0.005 &      \\ 
2457755.761   &   30.229   &    4.661 &   0.891   &    0.003 &   -20.150   &    2.680 &   -4.580   &    11.815 &   0.680   &    0.002 &      \\ 
2457761.605   &   35.904   &    1.162 &   0.903   &    0.012 &   17.051   &    5.987 &   -79.339   &    47.055 &   0.668   &    0.008 &      \\ 
2457802.576   &   38.515   &    1.380 &   0.895   &    0.002 &   -15.008   &    2.414 &   1.550   &    9.135 &   0.667   &    0.002 &      \\ 
2457806.389   &   41.947   &    3.338 &   0.896   &    0.002 &   0.799   &    1.004 &   -2.690   &    9.193 &   0.674   &    0.002 &      \\ 
2457808.567   &   26.849   &    2.715 &   0.908   &    0.002 &   -0.102   &    1.041 &   -5.160   &    8.195 &   0.669   &    0.002 &      \\ 
2457814.640   &   12.245   &    1.275 &   0.896   &    0.011 &   -6.001   &    5.268 &   -54.738   &    34.411 &   0.672   &    0.007 &      \\ 
2457815.710   &   10.287   &    1.251 &   0.901   &    0.003 &   -2.292   &    1.770 &   -17.069   &    14.029 &   0.670   &    0.002 &      \\ 
2457817.535   &   5.204   &    1.669 &   0.891   &    0.001 &   3.138   &    0.583 &   14.174   &    6.942 &   0.671   &    0.001 &      \\ 
2457818.543   &   0.904   &    1.575 &   0.899   &    0.002 &   0.736   &    0.844 &   4.075   &    6.815 &   0.676   &    0.001 &      \\ 
2457819.514   &   -4.258   &    1.314 &   0.899   &    0.003 &   -9.198   &    1.849 &   -1.330   &    12.426 &   0.675   &    0.002 &      \\ 
2457821.508   &   -8.726   &    1.522 &   0.896   &    0.002 &   -3.237   &    1.088 &   -23.956   &    8.546 &   0.678   &    0.002 &      \\ 
2457822.541   &   -10.641   &    1.044 &   0.891   &    0.001 &   0.521   &    0.897 &   5.802   &    6.424 &   0.678   &    0.001 &      \\ 
2457823.548   &   -15.454   &    1.360 &   0.884   &    0.001 &   -0.295   &    0.846&   4.447   &    7.825 &   0.678   &    0.001 &      \\ 
2457824.550   &   -18.304   &    0.969 &   0.883   &    0.001 &   -0.400   &    0.873&   -1.452   &    6.559 &   0.675   &    0.001 &      \\ 
2457828.530   &   -30.189   &    1.389 &   0.883   &    0.001 &   -1.616   &    0.943&   -1.198   &    6.584 &   0.674   &    0.001 &      \\ 
2457829.517   &   -32.157   &    1.531 &   0.879   &    0.001 &   1.640   &    0.959&   2.397   &    6.274 &   0.676   &    0.001 &      \\ 
2457830.523   &   -35.605   &    1.172 &   0.882   &    0.003 &   -9.729   &    1.278&   -14.638   &    11.396 &   0.675   &    0.002 &      \\ 
2457833.528   &   -30.515   &    1.182 &   0.879   &    0.003 &   -9.117   &    1.370&   -19.370   &    14.112 &   0.676   &    0.003 &      \\ 
2457834.669   &   -23.824   &    1.904 &   0.897   &    0.002 &   -2.177   &    0.873&   9.455   &    7.265 &   0.669   &    0.001 &      \\ 
2457848.448   &   29.178   &    1.376 &   0.894   &    0.001 &   -1.141   &    0.790&   12.161   &    6.387 &   0.678   &    0.001 &      \\ 
2457852.629   &   20.694   &    1.194 &   0.899   &    0.003 &   -10.150   &    1.671&   -0.692   &    14.780 &   0.676   &    0.002 &      \\ 
2457853.451   &   16.839   &    1.271 &   0.885   &    0.001 &   3.452   &    0.903&   -1.078   &    7.039 &   0.669   &    0.001 &      \\ 
2457855.514   &   10.445   &    1.223 &   0.898   &    0.002 &   -0.681   &    0.952&   8.580   &    6.513 &   0.674   &    0.001 &      \\ 
2457856.462   &   8.496   &    1.318 &   0.893   &    0.002 &   3.158   &    0.838&   -7.704   &    5.334 &   0.675   &    0.001 &      \\ 
2457857.450   &   5.788   &    1.025 &   0.890   &    0.001 &   3.555   &    0.841&   -10.142   &    7.150 &   0.674   &    0.001 &      \\ 
2457858.457   &   1.675   &    1.264 &   0.893   &    0.001 &   3.004   &    0.852&   -0.844   &    8.197 &   0.676   &    0.001 &      \\ 
2457859.473   &   -0.909   &    1.353 &   0.890   &    0.001 &   2.898   &    0.884&   3.355   &    7.111 &   0.667   &    0.001 &      \\ 
2457860.448   &   0.628   &    4.037 &   0.915   &    0.001 &   2.274   &    0.689&   4.717   &    6.327 &   0.677   &    0.001 &      \\ 
2457861.461   &   -6.803   &    1.183 &   0.884   &    0.001 &   1.607   &    0.695&   8.414   &    6.290 &   0.673   &    0.001 &      \\ 
2457862.496   &   -9.278   &    1.068 &   0.920   &    0.002 &   3.442   &    0.702&   13.066   &    8.350 &   0.678   &    0.001 &      \\ 
2457863.450   &   -9.142   &    3.177 &   0.855   &    0.010 &   5.980   &    5.380&   -36.896   &    42.941 &   0.673   &    0.007 &      \\ 
2457864.455   &   -13.755   &    1.748 &   0.888   &    0.001 &   2.935   &    0.741&   11.337   &    5.223 &   0.668   &    0.001 &      \\ 
2457866.443   &   -22.329   &    1.544 &   0.904   &    0.001 &   1.538   &    0.680&   6.350   &    6.204 &   0.676   &    0.001 &      \\ 
2457869.574   &   -31.669   &    2.664 &   0.890   &    0.007 &   4.756   &    4.607&   -34.628   &    27.503 &   0.663   &    0.005 &      \\ 
2457875.494   &   -37.861   &    1.183 &   0.898   &    0.003 &   -1.833   &    1.439&   4.954   &    13.346 &   0.671   &    0.002 &      \\ 
2457876.447   &   -29.784   &    1.357 &   0.895   &    0.001 &   3.148   &    0.724&   0.773   &    5.956 &   0.670   &    0.001 &      \\ 
2457877.399   &   -17.075   &    1.214 &   0.897   &    0.001 &   2.656   &    0.772&   -4.975   &    6.089 &   0.674   &    0.001 &      \\ 
2457880.400   &   20.788   &    1.407 &   0.897   &    0.002 &   1.270   &    1.116&   6.489   &    10.789 &   0.668   &    0.002 &      \\ 
2457881.390   &   24.470   &    1.825 &   0.886   &    0.002 &   5.761   &    0.868&   13.859   &    6.818 &   0.674   &    0.001 &      \\ 
2457882.435   &   31.290   &    1.269 &   0.902   &    0.001 &   5.993   &    0.917&   12.412   &    5.375 &   0.674   &    0.001 &      \\ 
2457883.425   &   31.224   &    1.603 &   0.900   &    0.002 &   9.400   &    0.790&   20.383   &    6.684 &   0.674   &    0.001 &      \\ 
2457886.501   &   29.274   &    1.948 &   0.902   &    0.001 &   7.731   &    0.817&   7.188   &    6.435 &   0.666   &    0.001 &      \\ 
2457887.494   &   35.670   &    1.789 &   0.899   &    0.001 &   7.625   &    0.816&   0.051   &    5.641 &   0.675   &    0.001 &      \\ 
2457888.437   &   32.948   &    1.291 &   0.896   &    0.001 &   8.470   &    0.867&   1.062   &    7.027 &   0.674   &    0.001 &      \\ 
2457889.402   &   29.249   &    1.284 &   0.887   &    0.002 &   9.181   &    1.433&   14.445   &    11.204 &   0.678   &    0.002 &      \\ 
2457890.475   &   27.853   &    1.302 &   0.897   &    0.002 &   3.880   &    1.013&   15.522   &    10.595 &   0.677   &    0.002 &      \\ 
2457891.471   &   21.825   &    1.619 &   0.889   &    0.003 &   4.552   &    1.286&   -18.835   &    11.685 &   0.672   &    0.002 &      \\  
  
\hline           
\end{tabular}

\end{table*}

\begin{table*}
\caption{CARMENES optical Doppler measurements and activity index measurements of GJ\,1148  } 
\label{table:CARM_VIS2} 

\centering  

\begin{tabular}{c c c c c c c c c c c c c c } 

\hline\hline    
\noalign{\vskip 0.5mm}

Epoch [JD] & RV [m\,s$^{-1}$] & $\sigma_{RV}$ [m\,s$^{-1}$]  &  H$\alpha$ & $\sigma_{H\alpha}$  &  dLW & $\sigma_{\rm dLW}$  &  CRX & $\sigma_{\rm CRX}$  & cairt & $\sigma_{\rm cairt}$  & \\  

\hline     
\noalign{\vskip 0.5mm}    

2457892.351   &   29.343   &    1.586 &   0.892   &    0.001 &   3.068   &    0.584&   2.076   &    7.214 &   0.673   &    0.001 &      \\ 
2457893.405   &   22.756   &    1.554 &   0.900   &    0.001 &   3.020   &    0.812&   -4.402   &    7.580 &   0.674   &    0.001 &      \\ 
2457894.429   &   18.392   &    1.239 &   0.888   &    0.001 &   1.120   &    0.694&   -0.531   &    7.263 &   0.668   &    0.001 &      \\ 
2457907.463   &   -19.222   &    1.284 &   0.889   &    0.001 &   -0.074   &    0.621&   -6.695   &    6.682 &   0.665   &    0.001 &      \\ 
2457914.434   &   -41.269   &    1.500 &   0.964   &    0.002 &   2.734   &    0.889&   -12.319   &    10.879 &   0.673   &    0.002 &      \\ 
2457921.464   &   19.493   &    1.326 &   0.885   &    0.002 &   -6.493   &    1.150&   -11.739   &    10.940 &   0.672   &    0.002 &      \\ 
2457931.419   &   25.190   &    2.301 &   0.931   &    0.001 &   0.196   &    0.692&   -13.134   &    5.632 &   0.677   &    0.001 &      \\ 
2457938.377   &   9.876   &    2.233 &   0.904   &    0.002 &   -2.306   &    1.247&   -6.690   &    8.876 &   0.676   &    0.001 &      \\ 
2457945.370   &   -14.280   &    1.422 &   0.906   &    0.001 &   1.395   &    1.091&   -1.648   &    7.533 &   0.670   &    0.001 &      \\ 
2457954.366   &   -45.774   &    1.876 &   0.909   &    0.002 &   -4.299   &    0.910&   0.390   &    7.885 &   0.676   &    0.001 &      \\ 
2457962.363   &   5.407   &    1.218 &   0.892   &    0.003 &   -17.897   &    2.189&   -27.132   &    14.522 &   0.677   &    0.002 &      \\ 
2457970.361   &   28.447   &    1.452 &   0.887   &    0.002 &   -10.692   &    1.810&   -15.832   &    6.093 &   0.667   &    0.001 &      \\ 
2458165.734   &   -32.856   &    1.760 &   0.893   &    0.002 &   -5.237   &    2.763&   -2.934   &    10.601 &   0.678   &    0.002 &      \\ 
2458172.579   &   34.563   &    1.662 &   0.896   &    0.002 &   -3.582   &    1.295&   12.421   &    12.907 &   0.676   &    0.002 &      \\ 
2458186.670   &   6.727   &    1.893 &   0.936   &    0.002 &   -2.205   &    1.108&   -3.476   &    9.360 &   0.680   &    0.001 &      \\ 
2458205.445   &   -37.733   &    1.177 &   0.901   &    0.002 &   -1.032   &    1.284&   -9.872   &    11.484 &   0.680   &    0.001 &      \\ 
2458212.470   &   32.105   &    1.563 &   0.911   &    0.002 &   -9.102   &    1.538&   -9.499   &    6.954 &   0.677   &    0.001 &      \\ 
2458225.597   &   15.118   &    1.276 &   0.894   &    0.002 &   -5.298   &    2.228&   -8.064   &    7.422 &   0.674   &    0.001 &      \\ 
2458599.510   &   -9.220   &    3.984 &   0.910   &    0.003 &   -3.332   &    1.531&   -31.015   &    11.867 &   0.675   &    0.002 &      \\ 
2458602.570   &   -14.322   &    1.741 &   0.933   &    0.002 &   3.704   &    0.746&   -13.568   &    6.446 &   0.672   &    0.001 &      \\ 
2458605.514   &   -25.749   &    1.265 &   0.928   &    0.004 &   -6.522   &    1.965&   5.337   &    13.854 &   0.678   &    0.003 &      \\ 
2458608.499   &   -37.237   &    1.307 &   0.899   &    0.001 &   1.897   &    0.786&   -3.168   &    7.024 &   0.674   &    0.001 &      \\ 
  
\hline           
\end{tabular}

\end{table*}

\begin{table}
\caption{CARMENES near-IR Doppler measurements for GJ\,1148} 
\label{tab:CARM_NIR} 

\centering  

\begin{tabular}{c c c } 

\hline\hline    
\noalign{\vskip 0.5mm}

Epoch [JD] & RV [m\,s$^{-1}$] & $\sigma_{RV}$ [m\,s$^{-1}$]   \\  

\hline     
\noalign{\vskip 0.5mm}    

2457754.710   &   -3.032   &    6.999      \\ 
2457761.605   &   30.494   &    11.595      \\ 
2457802.577   &   24.419   &    6.437      \\ 
2457806.389   &   19.870   &    21.382      \\ 
2457814.640   &   6.127   &    5.098      \\ 
2457815.709   &   0.209   &    6.261      \\ 
2457817.535   &   4.364   &    7.748      \\ 
2457818.546   &   -14.694   &    7.482      \\ 
2457819.516   &   -10.901   &    4.913      \\ 
2457821.508   &   -7.852   &    5.275      \\ 
2457822.540   &   -16.460   &    5.222      \\ 
2457823.549   &   -26.291   &    5.914      \\ 
2457824.553   &   -29.161   &    5.153      \\ 
2457828.530   &   -40.425   &    6.941      \\ 
2457829.519   &   -41.002   &    7.582      \\ 
2457830.524   &   -43.545   &    4.656      \\ 
2457833.527   &   -32.072   &    6.415      \\ 
2457834.671   &   -44.802   &    8.999      \\ 
2457848.447   &   27.418   &    5.415      \\ 
2457852.629   &   13.341   &    4.597      \\ 
2457853.451   &   11.756   &    5.522      \\ 
2457855.515   &   4.310   &    5.303      \\ 
2457856.460   &   12.351   &    10.617      \\ 
2457857.449   &   8.610   &    4.199      \\ 
2457858.457   &   -1.430   &    4.829      \\ 
2457859.471   &   0.000   &    5.351      \\ 
2457860.448   &   -21.713   &    15.841      \\ 
2457861.461   &   -3.867   &    4.358      \\ 
2457862.497   &   -7.161   &    6.096      \\ 
2457863.450   &   -30.657   &    11.949      \\ 
2457864.454   &   -32.249   &    7.966      \\ 
2457866.444   &   -22.153   &    5.543      \\ 
2457869.574   &   -45.875   &    17.141      \\ 
2457875.495   &   -41.333   &    5.491      \\ 
2457876.448   &   -27.119   &    5.711      \\ 
2457877.398   &   -19.532   &    4.108      \\ 
2457880.399   &   18.447   &    5.951      \\ 
2457881.391   &   10.319   &    7.016      \\

\hline           
\end{tabular}

\end{table}

\begin{table}
\caption{CARMENES near-IR Doppler measurements for GJ\,1148  } 
\label{tab:CARM_NIR2}  

\centering  

\begin{tabular}{c c c } 

\hline\hline    
\noalign{\vskip 0.5mm}

Epoch [JD] & RV [m\,s$^{-1}$] & $\sigma_{RV}$ [m\,s$^{-1}$]   \\  

\hline     
\noalign{\vskip 0.5mm}    

2457882.433   &   43.254   &    6.342      \\ 
2457883.427   &   32.059   &    5.848      \\ 
2457886.499   &   31.740   &    9.008      \\ 
2457887.491   &   25.641   &    6.372      \\ 
2457888.437   &   28.239   &    5.729      \\ 
2457889.401   &   20.735   &    7.268      \\ 
2457890.476   &   35.249   &    9.135      \\ 
2457891.470   &   15.146   &    6.185      \\ 
2457892.354   &   18.375   &    7.212      \\ 
2457893.405   &   24.157   &    6.362      \\ 
2457894.431   &   12.234   &    6.195      \\ 
2457907.463   &   -28.768   &    6.120      \\ 
2457914.433   &   -44.109   &    5.197      \\ 
2457921.462   &   11.233   &    6.030      \\ 
2457931.415   &   25.520   &    9.954      \\ 
2457938.378   &   22.890   &    9.100      \\ 
2457945.370   &   1.562   &    8.857      \\ 
2457954.366   &   -30.582   &    8.611      \\ 
2457962.363   &   7.001   &    12.502      \\ 
2457970.360   &   5.298   &    10.262      \\ 
2458165.734   &   -39.210   &    7.393      \\ 
2458172.579   &   33.807   &    12.316      \\ 
2458186.669   &   -4.074   &    9.982      \\ 
2458205.445   &   -38.023   &    5.123      \\ 
2458212.470   &   31.225   &    14.378      \\ 
2458225.598   &   18.548   &    5.504      \\ 
2458599.510   &   -28.057   &    16.599      \\ 
2458602.569   &   -13.256   &    9.232      \\ 
2458605.514   &   -27.589   &    5.967      \\ 
2458608.498   &   -38.473   &    5.744      \\

\hline           
\end{tabular}

\end{table}

\end{appendix}

\end{document}